\documentclass[%
reprint,
superscriptaddress,
nofootinbib
amsmath,amssymb,
aps,
floatfix,
nofootinbib
]{revtex4-2}

\usepackage{graphicx}
\usepackage{dcolumn}
\usepackage{bm}
\usepackage{acronym}
\usepackage{xspace}
\graphicspath{ {./figures/} }
\usepackage{color}
\usepackage{xstring}
\usepackage[table]{xcolor}
\usepackage{amsmath}
\usepackage{tabularx}
\usepackage[table]{xcolor}
\definecolor{lightgray}{gray}{0.9} 
\usepackage{tabularx}
\usepackage{multirow}
\usepackage{bigstrut}
\def\mystrut{\vrule height 9.3pt depth 3.1pt width 0pt}
\usepackage{booktabs}
\usepackage{orcidlink}

\newcommand{\figref}[1]{Fig.~\ref{#1}}

\newenvironment{event_table}{\setlength{\tabcolsep}{0pt}\setlength{\aboverulesep}{1pt}\setlength{\belowrulesep}{1pt}}{}

\begin{document}

\title{Performance of the low-latency GstLAL inspiral search towards LIGO, Virgo, and KAGRA’s fourth observing run}

\author{Becca Ewing \orcidlink{0000-0001-9178-5744}}
\email{rebecca.ewing@ligo.org}
\affiliation{Department of Physics, The Pennsylvania State University, University Park, PA 16802, USA}
\affiliation{Institute for Gravitation and the Cosmos, The Pennsylvania State University, University Park, PA 16802, USA}

\author{Rachael Huxford}
\affiliation{Department of Physics, The Pennsylvania State University, University Park, PA 16802, USA}
\affiliation{Institute for Gravitation and the Cosmos, The Pennsylvania State University, University Park, PA 16802, USA}

\author{Divya Singh \orcidlink{0000-0001-9675-4584}}
\affiliation{Department of Physics, The Pennsylvania State University, University Park, PA 16802, USA}
\affiliation{Institute for Gravitation and the Cosmos, The Pennsylvania State University, University Park, PA 16802, USA}

\author{Leo Tsukada  \orcidlink{0000-0003-0596-5648}}
\affiliation{Department of Physics, The Pennsylvania State University, University Park, PA 16802, USA}
\affiliation{Institute for Gravitation and the Cosmos, The Pennsylvania State University, University Park, PA 16802, USA}

\author{Chad Hanna}
\affiliation{Department of Physics, The Pennsylvania State University, University Park, PA 16802, USA}
\affiliation{Institute for Gravitation and the Cosmos, The Pennsylvania State University, University Park, PA 16802, USA}
\affiliation{Department of Astronomy and Astrophysics, The Pennsylvania State University, University Park, PA 16802, USA}
\affiliation{Institute for Computational and Data Sciences, The Pennsylvania State University, University Park, PA 16802, USA}

\author{Yun-Jing Huang \orcidlink{0000-0002-2952-8429}}
\affiliation{Department of Physics, The Pennsylvania State University, University Park, PA 16802, USA}
\affiliation{Institute for Gravitation and the Cosmos, The Pennsylvania State University, University Park, PA 16802, USA}

\author{Prathamesh Joshi \orcidlink{0000-0002-4148-4932}}
\affiliation{Department of Physics, The Pennsylvania State University, University Park, PA 16802, USA}
\affiliation{Institute for Gravitation and the Cosmos, The Pennsylvania State University, University Park, PA 16802, USA}

\author{Alvin K. Y. Li \orcidlink{0000-0001-6728-6523}}
\affiliation{LIGO Laboratory, California Institute of Technology, Pasadena, CA 91125, USA}

\author{Ryan Magee \orcidlink{0000-0001-9769-531X}}
\affiliation{LIGO Laboratory, California Institute of Technology, Pasadena, CA 91125, USA}

\author{Cody Messick}
\affiliation{MIT Kavli Institute for Astrophysics and Space Research, Massachusetts Institute of Technology, Cambridge, MA 02139, USA}

\author{Alex Pace}
\affiliation{Department of Physics, The Pennsylvania State University, University Park, PA 16802, USA}
\affiliation{Institute for Gravitation and the Cosmos, The Pennsylvania State University, University Park, PA 16802, USA}

\author{Anarya Ray \orcidlink{0000-0002-7322-4748}}
\affiliation{Leonard E.\ Parker Center for Gravitation, Cosmology, and Astrophysics, University of Wisconsin-Milwaukee, Milwaukee, WI 53201, USA}

\author{Surabhi Sachdev \orcidlink{0000-0002-0525-2317}}
\affiliation{School of Physics, Georgia Institute of Technology, Atlanta, GW 30332, USA}
\affiliation{Leonard E.\ Parker Center for Gravitation, Cosmology, and Astrophysics, University of Wisconsin-Milwaukee, Milwaukee, WI 53201, USA}

\author{Shio Sakon \orcidlink{0000-0002-5861-3024}}
\affiliation{Department of Physics, The Pennsylvania State University, University Park, PA 16802, USA}
\affiliation{Institute for Gravitation and the Cosmos, The Pennsylvania State University, University Park, PA 16802, USA}

\author{Ron Tapia}
\affiliation{Department of Physics, The Pennsylvania State University, University Park, PA 16802, USA}
\affiliation{Institute for Computational and Data Sciences, The Pennsylvania State University, University Park, PA 16802, USA}

\author{Shomik Adhicary}
\affiliation{Department of Physics, The Pennsylvania State University, University Park, PA 16802, USA}
\affiliation{Institute for Gravitation and the Cosmos, The Pennsylvania State University, University Park, PA 16802, USA}

\author{Pratyusava Baral \orcidlink{0000-0001-6308-211X}}
\affiliation{Leonard E.\ Parker Center for Gravitation, Cosmology, and Astrophysics, University of Wisconsin-Milwaukee, Milwaukee, WI 53201, USA}

\author{Amanda Baylor \orcidlink{0000-0003-0918-0864}}
\affiliation{Leonard E.\ Parker Center for Gravitation, Cosmology, and Astrophysics, University of Wisconsin-Milwaukee, Milwaukee, WI 53201, USA}

\author{Kipp Cannon \orcidlink{0000-0003-4068-6572}}
\affiliation{RESCEU, The University of Tokyo, Tokyo, 113-0033, Japan}

\author{Sarah Caudill}
\affiliation{Department of Physics, University of Massachusetts, Dartmouth, MA 02747, USA}
\affiliation{Center for Scientific Computing and Data Science Research, University of Massachusetts, Dartmouth, MA 02747, USA}

\author{Sushant Sharma Chaudhary}
\affiliation{Institute of Multi-messenger Astrophysics and Cosmology, Missouri University of Science and Technology, Physics Building, 1315 N. Pine St., Rolla, MO 65409, USA}

\author{Michael W. Coughlin \orcidlink{0000-0002-8262-2924}}
\affiliation{School of Physics and Astronomy, University of Minnesota,
Minneapolis, Minnesota 55455, USA}
\thanks{MWC acknowledges support from the National Science Foundation with grant numbers PHY-2010970 and OAC-2117997.}

\author{Bryce Cousins \orcidlink{0000-0002-7026-1340}}
\affiliation{Department of Physics, University of Illinois, Urbana, IL 61801 USA}
\affiliation{Department of Physics, The Pennsylvania State University, University Park, PA 16802, USA}
\affiliation{Institute for Gravitation and the Cosmos, The Pennsylvania State University, University Park, PA 16802, USA}

\author{Jolien D. E. Creighton \orcidlink{0000-0003-3600-2406}}
\affiliation{Leonard E.\ Parker Center for Gravitation, Cosmology, and Astrophysics, University of Wisconsin-Milwaukee, Milwaukee, WI 53201, USA}

\author{Reed Essick}
\affiliation{Canadian Institute for Theoretical Astrophysics, 60 St George St, Toronto, ON M5S 3H8}
\thanks{R.E is supported by the Natural Sciences \& Engineering Research Council of Canada (NSERC).}

\author{Heather Fong}
\affiliation{RESCEU, The University of Tokyo, Tokyo, 113-0033, Japan}
\affiliation{Graduate School of Science, The University of Tokyo, Tokyo 113-0033, Japan}

\author{Richard N. George \orcidlink{0000-0002-7797-7683}}
\affiliation{Center for Gravitational Physics, University of Texas at Austin, Austin, TX 78712, USA}

\author{Patrick Godwin}
\affiliation{LIGO Laboratory, California Institute of Technology, MS 100-36, Pasadena, California 91125, USA}
\affiliation{Department of Physics, The Pennsylvania State University, University Park, PA 16802, USA}
\affiliation{Institute for Gravitation and the Cosmos, The Pennsylvania State University, University Park, PA 16802, USA}

\author{Reiko Harada}
\affiliation{RESCEU, The University of Tokyo, Tokyo, 113-0033, Japan}
\affiliation{Graduate School of Science, The University of Tokyo, Tokyo 113-0033, Japan}

\author{James Kennington \orcidlink{0000-0002-6899-3833}}
\affiliation{Department of Physics, The Pennsylvania State University, University Park, PA 16802, USA}
\affiliation{Institute for Gravitation and the Cosmos, The Pennsylvania State University, University Park, PA 16802, USA}

\author{Soichiro Kuwahara}
\affiliation{RESCEU, The University of Tokyo, Tokyo, 113-0033, Japan}
\affiliation{Graduate School of Science, The University of Tokyo, Tokyo 113-0033, Japan}

\author{Duncan Meacher \orcidlink{0000-0001-5882-0368}}
\affiliation{Leonard E.\ Parker Center for Gravitation, Cosmology, and Astrophysics, University of Wisconsin-Milwaukee, Milwaukee, WI 53201, USA}

\author{Soichiro Morisaki \orcidlink{0000-0002-8445-6747}}
\affiliation{Institute for Cosmic Ray Research, The University of Tokyo, 5-1-5 Kashiwanoha, Kashiwa, Chiba 277-8582, Japan}
\affiliation{Leonard E.\ Parker Center for Gravitation, Cosmology, and Astrophysics, University of Wisconsin-Milwaukee, Milwaukee, WI 53201, USA}

\author{Debnandini Mukherjee  \orcidlink{0000-0001-7335-9418}}
\affiliation{NASA Marshall Space Flight Center, Huntsville, AL 35811, USA}
\affiliation{Center for Space Plasma and Aeronomic Research, University of Alabama in Huntsville, Huntsville, AL 35899, USA}

\author{Wanting Niu}
\affiliation{Department of Physics, The Pennsylvania State University, University Park, PA 16802, USA}
\affiliation{Institute for Gravitation and the Cosmos, The Pennsylvania State University, University Park, PA 16802, USA}

\author{Cort Posnansky}
\affiliation{Department of Physics, The Pennsylvania State University, University Park, PA 16802, USA}
\affiliation{Institute for Gravitation and the Cosmos, The Pennsylvania State University, University Park, PA 16802, USA}

\author{Andrew Toivonen \orcidlink{0009-0008-9546-2035}}
\affiliation{School of Physics and Astronomy, University of Minnesota,
Minneapolis, Minnesota 55455, USA}
\thanks{MWC acknowledges support from the National Science Foundation with grant numbers PHY-2010970 and OAC-2117997.}

\author{Takuya Tsutsui \orcidlink{0000-0002-2909-0471}}
\affiliation{RESCEU, The University of Tokyo, Tokyo, 113-0033, Japan}

\author{Koh Ueno \orcidlink{0000-0003-3227-6055}}
\affiliation{RESCEU, The University of Tokyo, Tokyo, 113-0033, Japan}

\author{Aaron Viets \orcidlink{0000-0002-4241-1428}}
\affiliation{Concordia University Wisconsin, Mequon, WI 53097, USA}

\author{Leslie Wade}
\affiliation{Department of Physics, Hayes Hall, Kenyon College, Gambier, Ohio 43022, USA}

\author{Madeline Wade \orcidlink{0000-0002-5703-4469}}
\affiliation{Department of Physics, Hayes Hall, Kenyon College, Gambier, Ohio 43022, USA}

\author{Gaurav Waratkar\orcidlink{0000-0003-3630-9440}}
\affiliation{Department of Physics, IIT Bombay, Powai, Mumbai 400076, India}

\date{May 9, 2023}

\begin{abstract}
GstLAL is a stream-based matched-filtering search pipeline aiming at the prompt discovery of gravitational waves from compact binary coalescences such as the mergers of black holes and neutron stars. 
Over the past three observation runs by the LIGO, Virgo, and KAGRA (LVK) collaboration, the GstLAL search pipeline has participated in several tens of gravitational wave discoveries. 
The fourth observing run (O4) is set to begin in May 2023 and is expected to see the discovery of many new and interesting gravitational wave signals which will inform our understanding of astrophysics and cosmology.
We describe the current configuration of the GstLAL low-latency search and show its readiness for the upcoming observation run by presenting its performance on a mock data challenge. 
The mock data challenge includes $40$ days of LIGO Hanford, LIGO Livingston, and Virgo strain data along with an injection campaign in order to fully characterize the performance of the search. 
We find an improved performance in terms of detection rate and significance estimation as compared to that observed in the O3 online analysis. 
The improvements are attributed to several incremental advances in the likelihood ratio ranking statistic computation and the method of background estimation.
\end{abstract}


\maketitle


\acrodef{LVK}[LVK]{LIGO Scientific, Virgo and KAGRA}
\acrodef{LDG}[LDG]{LIGO Data Grid}

\acrodef{L1}[L]{LIGO Livingston}
\acrodef{H1}[H]{LIGO Hanford}
\acrodef{V1}[V]{Virgo}

\acrodef{O4}[O4]{fourth observing run}
\acrodef{O3}[O3]{third observing run}
\acrodef{O2}[O2]{second observing run}
\acrodef{O1}[O1]{first observing run}

\acrodef{GWTC}[GWTC-3]{Gravitational Wave Transient Catalog}

\acrodef{GPS}[GPS]{Global Positioning System}
\acrodef{GCN}[GCN]{Gamma-ray Coordinate Network}

\acrodef{BH}[BH]{black hole}
\acrodef{IMBH}[IMBH]{intermediate mass black hole}
\acrodef{BBH}[BBH]{binary black hole}
\acrodefplural{BBH}[BBHs]{binary black holes}
\acrodef{BNS}[BNS]{binary neutron star}
\acrodef{NS}[NS]{neutron star}
\acrodef{NSBH}[NSBH]{neutron star--black hole binary}
\acrodefplural{NSBH}[NSBHs]{neutron star--black hole binaries}
\acrodef{CBC}[CBC]{compact binary coalescence}
\acrodef{GW}[GW]{gravitational wave}
\acrodef{SSM}[SSM]{sub-solar mass black hole}

\acrodef{SNR}[SNR]{signal-to-noise ratio}
\acrodefplural{SNR}[SNRs]{signal-to-noise ratios}
\acrodef{FAR}[FAR]{false alarm rate}
\acrodefplural{FAR}[FARs]{false alarm rates}
\acrodef{IFAR}[IFAR]{inverse false alarm rate}
\acrodef{PSD}[PSD]{power spectral density}
\acrodefplural{PSD}[PSDs]{power spectral densities}
\acrodef{SVD}[SVD]{singular value decomposition}
\acrodef{FIR}[FIR]{finite impulse response}
\acrodef{MDC}[MDC]{Mock Data Challenge}
\acrodef{LLOID}[LLOID]{Low-Latency Online Inspiral Detection}
\acrodef{PN}[PN]{Post-Newtonian}
\acrodef{FFT}[FFT]{fast Fourier transform}

\acrodef{GRACEDB}[GraceDB]{Gravitational wave Candidate Event Database}
\acrodef{GWLTS}[\texttt{gw-lts}]{Gravitational Wave Low-latency Test Suite}
\acrodef{IDQ}[iDQ]{integrated Data Quality}

\newcommand\hmm[1]{\ifnum\ifhmode\spacefactor\else2000\fi>1500 \uppercase{#1}\else#1\fi}

\newcommand{\GSTLAL}{GstLAL\xspace}
\newcommand{\IGWNALERT}{\texttt{igwn-alert}\xspace}

\newcommand{\MDCSTART}{5 Jan. 2020 15:59:42}
\newcommand{\MDCEND}{14 Feb. 2020 15:59:42}

\newcommand{\NUMSVDBANKS}{\ensuremath{\sim 1000}}
\newcommand{\TEMPLATESPERSUBBANK}{\ensuremath{\sim 500}}
\newcommand{\NUMSUBBANKSPERSVD}{\ensuremath{2}}
\newcommand{\SVDTOLERANCE}{\ensuremath{99.999\%}}
\newcommand{\PSDFFTLENGTH}{\ensuremath{4~\mathrm{seconds}}}
\newcommand{\FRAMELENGTH}{\ensuremath{1~\mathrm{second}}}
\newcommand{\BUFFERBLOCKSIZE}{\ensuremath{4096~\mathrm{bytes}}}
\newcommand{\FIRSTRIDE}{\ensuremath{0.25~\mathrm{seconds}}}
\newcommand{\TRIGGERSNRTHRESHOLD}{\ensuremath{4.0}}
\newcommand{\COINCTHRESHOLD}{\ensuremath{0.005~\mathrm{seconds}}}
\newcommand{\HTGATETHRESHOLDMIN}{\ensuremath{15.0}}
\newcommand{\HTGATETHRESHOLDMAX}{\ensuremath{100.0}}
\newcommand{\HTGATEMCHIRPMIN}{\ensuremath{0.8}}
\newcommand{\HTGATEMCHIRPMAX}{\ensuremath{45.0}}
\newcommand{\HTGATEMIN}{\ensuremath{\sim 15}}
\newcommand{\HTGATEMAX}{\ensuremath{\sim 325}}
\newcommand{\LRSNAPSHOT}{\ensuremath{4}}
\newcommand{\LRCOMPRESSION}{\ensuremath{0.003}}
\newcommand{\FARTRIALSFACTOR}{\ensuremath{2}}
\newcommand{\UPLOADCADENCE}{\ensuremath{4}}
\newcommand{\UPLOADCADENCEMDCTWELVE}{\ensuremath{2}}
\newcommand{\UPLOADDT}{\ensuremath{0.2}}
\newcommand{\SINGLESPENALTYMDCELEVEN}{\ensuremath{12}}
\newcommand{\SINGLESPENALTYOFOUR}{\ensuremath{13}}
\newcommand{\XISQMISMATCHRANGE}{\ensuremath{0.1-10\%}}

\newcommand{\TESTSUITECOINCWINDOW}{\ensuremath{\pm 1}}
\newcommand{\INJSNRFLOW}{\ensuremath{10.0}}
\newcommand{\INJSNRFHI}{\ensuremath{1600.0}}

\newcommand{\VT}{\ensuremath{\langle VT \rangle}}
\newcommand{\SPINZ}{\ensuremath{s_{i,z}}}
\newcommand{\CHIP}{\ensuremath{\chi_p}}
\newcommand{\MCHIRP}{\ensuremath{\mathcal{M}_c}\xspace}
\newcommand{\CHIEFF}{\ensuremath{\chi_{\mathrm{eff}}}}
\newcommand{\PASTRO}{\ensuremath{p(\mathrm{astro})}}
\newcommand{\MSUN}{\ensuremath{M_{\odot}}}
\newcommand{\TEND}{\ensuremath{t_{\mathrm{end}}}}

\newcommand{\TOTALINJECTIONS}{$5\times10^4$}
\newcommand{\BNSMAXZ}{\ensuremath{0.15}}
\newcommand{\NSBHMAXZ}{\ensuremath{0.25}}
\newcommand{\BBHMAXZ}{\ensuremath{1.9}}
\newcommand{\MDCDURATION}{\ensuremath{3.456\times10^6}}
\newcommand{\INJECTIONSPACING}{\ensuremath{\sim40}}

\newcommand{\DECISIVESNRTHRESH}{\ensuremath{8.0}}
\newcommand{\NETWORKSNRTHRESH}{\ensuremath{10.0}}

\newcommand{\ALLABOVEDECSNRTHRESH}{\ensuremath{1553}}
\newcommand{\ALLINBANKABOVEDECSNRTHRESH}{\ensuremath{1487}}
\newcommand{\BBHINBANKABOVEDECSNRTHRESH}{\ensuremath{624}}
\newcommand{\BNSINBANKABOVEDECSNRTHRESH}{\ensuremath{483}}
\newcommand{\NSBHINBANKABOVEDECSNRTHRESH}{\ensuremath{380}}

\newcommand{\HIGHFARTHRESH}{$1$ per hour}
\newcommand{\LOWFARTHRESH}{$2$ per day}
\newcommand{\ONEPERHOUR}{\ensuremath{2.78\times10^{-4}~\mathrm{Hz}}}
\newcommand{\TWOPERDAY}{\ensuremath{2.31\times10^{-5}~\mathrm{Hz}}}
\newcommand{\ONEPERMONTH}{\ensuremath{3.85\times10^{-7}~\mathrm{Hz}}}
\newcommand{\TWOPERYEAR}{\ensuremath{3.16\times10^{-8}~\mathrm{Hz}}}

\newcommand{\ALLINBANKEFFICIENCY}[1]{%
	\IfEqCase{#1}{%
		{ONEPERHOUR}{\ensuremath{0.87}}%
		{TWOPERDAY}{\ensuremath{0.84}}%
		{ONEPERMONTH}{\ensuremath{0.77}}%
		{TWOPERYEAR}{\ensuremath{0.72}}%
	}[\PackageError{ALLINBANKEFFICIENCY}{Undefined option: #1}{}]
}%

\newcommand{\BBHINBANKEFFICIENCY}[1]{%
	\IfEqCase{#1}{%
		{ONEPERHOUR}{\ensuremath{0.87}}%
		{TWOPERDAY}{\ensuremath{0.84}}%
		{ONEPERMONTH}{\ensuremath{0.75}}%
		{TWOPERYEAR}{\ensuremath{0.69}}%
	}[\PackageError{BBHINBANKEFFICIENCY}{Undefined option: #1}{}]
}%

\newcommand{\BNSINBANKEFFICIENCY}[1]{%
	\IfEqCase{#1}{%
		{ONEPERHOUR}{\ensuremath{0.95}}%
		{TWOPERDAY}{\ensuremath{0.94}}%
		{ONEPERMONTH}{\ensuremath{0.89}}%
		{TWOPERYEAR}{\ensuremath{0.86}}%
	}[\PackageError{BNSINBANKEFFICIENCY}{Undefined option: #1}{}]
}%

\newcommand{\NSBHINBANKEFFICIENCY}[1]{%
	\IfEqCase{#1}{%
		{ONEPERHOUR}{\ensuremath{0.77}}%
		{TWOPERDAY}{\ensuremath{0.71}}%
		{ONEPERMONTH}{\ensuremath{0.65}}%
		{TWOPERYEAR}{\ensuremath{0.62}}%
	}[\PackageError{NSBHINBANKEFFICIENCY}{Undefined option: #1}{}]
}%

\newcommand{\ALLABOVEDECSNRTHRESHMDCTWELVE}{\ensuremath{662}}
\newcommand{\ALLINBANKABOVEDECSNRTHRESHMDCTWELVE}{\ensuremath{630}}
\newcommand{\BBHINBANKABOVEDECSNRTHRESHMDCTWELVE}{\ensuremath{249}}
\newcommand{\BNSINBANKABOVEDECSNRTHRESHMDCTWELVE}{\ensuremath{209}}
\newcommand{\NSBHINBANKABOVEDECSNRTHRESHMDCTWELVE}{\ensuremath{172}}

\newcommand{\ALLINBANKEFFICIENCYMDCTWELVE}[1]{%
	\IfEqCase{#1}{%
		{ONEPERHOUR}{\ensuremath{0.87}}%
		{TWOPERDAY}{\ensuremath{0.86}}%
		{ONEPERMONTH}{\ensuremath{0.82}}%
		{TWOPERYEAR}{\ensuremath{0.80}}%
	}[\PackageError{ALLINBANKEFFICIENCYMDCTWELVE}{Undefined option: #1}{}]
}%

\newcommand{\BBHINBANKEFFICIENCYMDCTWELVE}[1]{%
	\IfEqCase{#1}{%
		{ONEPERHOUR}{\ensuremath{0.92}}%
		{TWOPERDAY}{\ensuremath{0.90}}%
		{ONEPERMONTH}{\ensuremath{0.86}}%
		{TWOPERYEAR}{\ensuremath{0.85}}%
	}[\PackageError{BBHINBANKEFFICIENCYMDCTWELVE}{Undefined option: #1}{}]
}%

\newcommand{\BNSINBANKEFFICIENCYMDCTWELVE}[1]{%
	\IfEqCase{#1}{%
		{ONEPERHOUR}{\ensuremath{0.95}}%
		{TWOPERDAY}{\ensuremath{0.93}}%
		{ONEPERMONTH}{\ensuremath{0.91}}%
		{TWOPERYEAR}{\ensuremath{0.88}}%
	}[\PackageError{BNSINBANKEFFICIENCYMDCTWELVE}{Undefined option: #1}{}]
}%

\newcommand{\NSBHINBANKEFFICIENCYMDCTWELVE}[1]{%
	\IfEqCase{#1}{%
		{ONEPERHOUR}{\ensuremath{0.74}}%
		{TWOPERDAY}{\ensuremath{0.72}}%
		{ONEPERMONTH}{\ensuremath{0.67}}%
		{TWOPERYEAR}{\ensuremath{0.66}}%
	}[\PackageError{NSBHINBANKEFFICIENCYMDCTWELVE}{Undefined option: #1}{}]
}%

\newcommand{\MEAN}[1]{%
	\IfEqCase{#1}{%
		{MASSRATIO}{\ensuremath{1.39}}%
		{MCHIRP}{\ensuremath{0.15}}%
		{SPIN1Z}{\ensuremath{7.27}}%
		{SPIN2Z}{\ensuremath{2.81}}%
		{CHIEFF}{\ensuremath{5.77}}%
		{ENDTIME}{\ensuremath{6.23}}%
	}[\PackageError{MEAN}{Undefined option: #1}{}]
}%

\newcommand{\STDEV}[1]{%
	\IfEqCase{#1}{%
		{MASSRATIO}{\ensuremath{2.86}}%
		{MCHIRP}{\ensuremath{0.45}}%
		{SPIN1Z}{\ensuremath{285}}%
		{SPIN2Z}{\ensuremath{183}}%
		{CHIEFF}{\ensuremath{252}}%
		{ENDTIME}{\ensuremath{30.22}}%
	}[\PackageError{STDEV}{Undefined option: #1}{}]
}%

\newcommand{\BNSMCHIRPMEAN}{\ensuremath{2.06\times10^{-4}}}
\newcommand{\BNSMCHIRPSTDEV}{\ensuremath{8.33\times10^{-4}}}

\newcommand{\NSBHMCHIRPMEAN}{\ensuremath{-2.14\times10^{-4}}}
\newcommand{\NSBHMCHIRPSTDEV}{\ensuremath{6.26\times10^{-3}}}

\newcommand{\BBHMCHIRPMEAN}{\ensuremath{1.54\times10^{-1}}}
\newcommand{\BBHMCHIRPSTDEV}{\ensuremath{4.53\times10^{-1}}}

\newcommand{\BNSENDTIMEMEAN}{\ensuremath{-0.90}}
\newcommand{\BNSENDTIMESTDEV}{\ensuremath{18.0}}

\newcommand{\NSBHENDTIMEMEAN}{\ensuremath{18.7}}
\newcommand{\NSBHENDTIMESTDEV}{\ensuremath{59.3}}

\newcommand{\BBHENDTIMEMEAN}{\ensuremath{6.03}}
\newcommand{\BBHENDTIMESTDEV}{\ensuremath{11.3}}

\newcommand{\QFIFTY}[1]{%
	\IfEqCase{#1}{%
		{MASSRATIO}{\ensuremath{0.45}}%
		{MCHIRP}{\ensuremath{0.007}}%
		{SPIN1Z}{\ensuremath{1.24}}%
		{SPIN2Z}{\ensuremath{1.72}}%
		{CHIEFF}{\ensuremath{1.34}}%
		{ENDTIME}{\ensuremath{3.8}}%
	}[\PackageError{QFIFTY}{Undefined option: #1}{}]
}%

\newcommand{\QSEVENTYFIVE}[1]{%
	\IfEqCase{#1}{%
		{MASSRATIO}{\ensuremath{1.67}}%
		{MCHIRP}{\ensuremath{0.33}}%
		{SPIN1Z}{\ensuremath{3.82}}%
		{SPIN2Z}{\ensuremath{5.33}}%
		{CHIEFF}{\ensuremath{3.71}}%
		{ENDTIME}{\ensuremath{9.78}}%
	}[\PackageError{QSEVENTYFIVE}{Undefined option: #1}{}]
}%

\newcommand{\QNINETY}[1]{%
	\IfEqCase{#1}{%
		{MASSRATIO}{\ensuremath{4.97}}%
		{MCHIRP}{\ensuremath{0.73}}%
		{SPIN1Z}{\ensuremath{13.8}}%
		{SPIN2Z}{\ensuremath{17.5}}%
		{CHIEFF}{\ensuremath{10.8}}%
		{ENDTIME}{\ensuremath{25.6}}%
	}[\PackageError{QNINETY}{Undefined option: #1}{}]
}%

\newcommand{\GPCYRS}{\ensuremath{\mathrm{Gpc}^3\mathrm{yrs}}}
\newcommand{\INJECTEDVT}[1]{%
	\IfEqCase{#1}{%
		{BNS}{\ensuremath{1.08\times10^{-1}}}%
		{NSBH}{\ensuremath{4.34\times10^{-1}}}%
		{BBH}{\ensuremath{29.1}}%
	}[\PackageError{INJECTEDVT}{Undefined option: #1}{}]
}%

\newcommand{\VTTWOPERDAY}[1]{%
	\IfEqCase{#1}{%
		{BNS}{\ensuremath{3.49\times10^{-4}}}%
		{NSBH}{\ensuremath{8.08\times10^{-4}}}%
		{BBH}{\ensuremath{1.23\times10^{-1}}}%
	}[\PackageError{VTTWOPERDAY}{Undefined option: #1}{}]
}%

\newcommand{\VTNETSNR}[1]{%
	\IfEqCase{#1}{%
		{BNS}{\ensuremath{4.41\times10^{-4}}}%
		{NSBH}{\ensuremath{1.59\times10^{-3}}}%
		{BBH}{\ensuremath{1.52\times10^{-1}}}%
	}[\PackageError{VTNETSNR}{Undefined option: #1}{}]
}%

\newcommand{\VTDECSNR}[1]{%
	\IfEqCase{#1}{%
		{BNS}{\ensuremath{1.47\times10^{-4}}}%
		{NSBH}{\ensuremath{4.98\times10^{-3}}}%
		{BBH}{\ensuremath{5.46\times10^{-2}}}%
	}[\PackageError{VTDECSNR}{Undefined option: #1}{}]
}%


\newcommand{\SEARCHEDAREAQFIFTY}[1]{%
	\IfEqCase{#1}{%
		{ALL}{\ensuremath{271}}%
		{TRIPLE}{\ensuremath{31.9}}%
		{DOUBLE}{\ensuremath{301}}%
		{SINGLE}{\ensuremath{3150}}%
	}[\PackageError{SEARCHEDAREAQFIFTY}{Undefined option: #1}{}]
}%

\newcommand{\SEARCHEDAREAQSEVENTYFIVE}[1]{%
	\IfEqCase{#1}{%
		{ALL}{\ensuremath{1080}}%
		{TRIPLE}{\ensuremath{140}}%
		{DOUBLE}{\ensuremath{893}}%
		{SINGLE}{\ensuremath{10,400}}%
	}[\PackageError{SEARCHEDAREAQSEVENTYFIVE}{Undefined option: #1}{}]
}%

\newcommand{\SEARCHEDAREAQNINETY}[1]{%
	\IfEqCase{#1}{%
		{ALL}{\ensuremath{3910}}%
		{TRIPLE}{\ensuremath{357}}%
		{DOUBLE}{\ensuremath{2470}}%
		{SINGLE}{\ensuremath{18,400}}%
	}[\PackageError{SEARCHEDAREAQNINETY}{Undefined option: #1}{}]
}%

\newcommand{\SEARCHEDPROBQFIFTY}[1]{%
	\IfEqCase{#1}{%
		{ALL}{\ensuremath{0.53}}%
		{TRIPLE}{\ensuremath{0.58}}%
		{DOUBLE}{\ensuremath{0.52}}%
		{SINGLE}{\ensuremath{0.59}}%
	}[\PackageError{SEARCHEDPROBQFIFTY}{Undefined option: #1}{}]
}%

\newcommand{\SEARCHEDPROBQSEVENTYFIVE}[1]{%
	\IfEqCase{#1}{%
		{ALL}{\ensuremath{0.79}}%
		{TRIPLE}{\ensuremath{0.84}}%
		{DOUBLE}{\ensuremath{0.77}}%
		{SINGLE}{\ensuremath{0.78}}%
	}[\PackageError{SEARCHEDPROBQSEVENTYFIVE}{Undefined option: #1}{}]
}%

\newcommand{\SEARCHEDPROBQNINETY}[1]{%
	\IfEqCase{#1}{%
		{ALL}{\ensuremath{0.93}}%
		{TRIPLE}{\ensuremath{0.96}}%
		{DOUBLE}{\ensuremath{0.92}}%
		{SINGLE}{\ensuremath{0.92}}%
	}[\PackageError{SEARCHEDPROBQNINETY}{Undefined option: #1}{}]
}%

\newcommand{\BNSTOBNS}{\ensuremath{90.3\%}}
\newcommand{\BNSTONSBH}{\ensuremath{9.7\%}}

\newcommand{\NSBHTONSBH}{\ensuremath{64.1\%}}
\newcommand{\NSBHTOBBH}{\ensuremath{33.8\%}}
\newcommand{\NSBHTOBNS}{\ensuremath{2.10\%}}

\newcommand{\BBHTOBBH}{\ensuremath{100\%}}

\newcommand{\TERRTOTERR}{\ensuremath{2.60\%}}
\newcommand{\TERRTOBBH}{\ensuremath{68.8\%}}
\newcommand{\TERRTONSBH}{\ensuremath{19.5\%}}
\newcommand{\TERRTOBNS}{\ensuremath{9.10\%}}

\newcommand{\BNSTOBNSMDCTWELVE}{\ensuremath{79.8\%}}
\newcommand{\BNSTONSBHMDCTWELVE}{\ensuremath{20.2\%}}

\newcommand{\NSBHTONSBHMDCTWELVE}{\ensuremath{92.1\%}}
\newcommand{\NSBHTOBBHMDCTWELVE}{\ensuremath{6.83\%}}
\newcommand{\NSBHTOBNSMDCTWELVE}{\ensuremath{1.02\%}}

\newcommand{\BBHTOBBHMDCTWELVE}{\ensuremath{99.5\%}}
\newcommand{\BBHTONSBHMDCTWELVE}{\ensuremath{0.05\%}}

\newcommand{\TERRTOBBHMDCTWELVE}{\ensuremath{76.2\%}}
\newcommand{\TERRTONSBHMDCTWELVE}{\ensuremath{23.8\%}}

\newcommand{\OTHREEOPA}{\ensuremath{1.2}} 

\newcommand{\MDCGWIFOS}[1]{%
	\IfEqCase{#1}{%
		{GW200112}{L1}%
		{GW200115}{H1L1}%
		{GW200128}{H1L1}%
		{GW200129}{H1L1V1}%
		{GW200202}{H1L1}%
		{GW200208q}{H1L1}%
		{GW200208am}{H1L1}%
		{GW200209}{H1L1}%
		{GW200210}{H1L1}%
	}[\PackageError{MDCGWIFOS}{Undefined option: #1}{}]
}%

\newcommand{\MDCGWSNR}[1]{%
	\IfEqCase{#1}{%
		{GW200112}{\ensuremath{18.46}}%
		{GW200115}{\ensuremath{11.48}}%
		{GW200128}{\ensuremath{9.98}}%
		{GW200129}{\ensuremath{26.30}}%
		{GW200202}{\ensuremath{11.09}}%
		{GW200208q}{\ensuremath{10.56}}%
		{GW200208am}{\ensuremath{8.00}}%
		{GW200209}{\ensuremath{9.96}}%
		{GW200210}{\ensuremath{9.28}}%
	}[\PackageError{MDCGWSNR}{Undefined option: #1}{}]
}%

\newcommand{\MDCGWFAR}[1]{%
	\IfEqCase{#1}{%
		{GW200112}{\ensuremath{1.01\times10^{-7}}}%
		{GW200115}{\ensuremath{2.55\times10^{-4}}}%
		{GW200128}{\ensuremath{1.44\times10^{-4}}}%
		{GW200129}{\ensuremath{1.78\times10^{-17}}}%
		{GW200202}{\ensuremath{1.69\times10^{-2}}}%
		{GW200208q}{\ensuremath{4.92\times10^{-5}}}%
		{GW200208am}{\ensuremath{2.02\times10^{3}}}%
		{GW200209}{\ensuremath{1.20}}%
		{GW200210}{\ensuremath{3.64\times10^{3}}}%
	}[\PackageError{MDCGWFAR}{Undefined option: #1}{}]
}%

\newcommand{\MDCGWPASTRO}[1]{%
	\IfEqCase{#1}{%
		{GW200112}{\ensuremath{>0.99}}%
		{GW200115}{\ensuremath{>0.99}}%
		{GW200128}{\ensuremath{>0.99}}%
		{GW200129}{\ensuremath{>0.99}}%
		{GW200202}{\ensuremath{>0.99}}%
		{GW200208q}{\ensuremath{>0.99}}%
		{GW200208am}{\ensuremath{0.48}}%
		{GW200209}{\ensuremath{>0.99}}%
		{GW200210}{0.27}%
	}[\PackageError{MDCGWPASTRO}{Undefined option: #1}{}]
}%

\newcommand{\MDCGWMCHIRP}[1]{%
	\IfEqCase{#1}{%
		{GW200112}{\ensuremath{33.37~M_{\odot}}}%
		{GW200115}{\ensuremath{2.58~M_{\odot}}}%
		{GW200128}{\ensuremath{50.74~M_{\odot}}}%
		{GW200129}{\ensuremath{30.66~M_{\odot}}}%
		{GW200202}{\ensuremath{8.15~M_{\odot}}}%
		{GW200208q}{\ensuremath{34.50~M_{\odot}}}%
		{GW200208am}{\ensuremath{66.59~M_{\odot}}}%
		{GW200209}{\ensuremath{39.45~M_{\odot}}}%
		{GW200210}{\ensuremath{7.89~M_{\odot}}}%
	}[\PackageError{MDCGWMCHIRP}{Undefined option: #1}{}]
}%

\newcommand{\OTHREEGWIFOS}[1]{%
	\IfEqCase{#1}{%
		{GW200112}{L1}%
		{GW200115}{H1L1}%
		{GW200128}{--}%
		{GW200129}{H1L1V1}%
		{GW200202}{--}%
		{GW200208q}{--}%
		{GW200208am}{--}%
		{GW200209}{--}%
		{GW200210}{--}%
	}[\PackageError{OTHREEGWIFOS}{Undefined option: #1}{}]
}%

\newcommand{\OTHREEGWSNR}[1]{%
	\IfEqCase{#1}{%
		{GW200112}{\ensuremath{18.79}}%
		{GW200115}{\ensuremath{11.42}}%
		{GW200128}{--}%
		{GW200129}{\ensuremath{26.61}}%
		{GW200202}{--}%
		{GW200208q}{--}%
		{GW200208am}{--}%
		{GW200209}{--}%
		{GW200210}{--}%
	}[\PackageError{OTHREEGWSNR}{Undefined option: #1}{}]
}%

\newcommand{\OTHREEGWFAR}[1]{%
	\IfEqCase{#1}{%
		{GW200112}{\ensuremath{4.05\times10^{-4}}}%
		{GW200115}{\ensuremath{6.61\times10^{-4}}}%
		{GW200128}{\ensuremath{> \OTHREEOPA{}}}%
		{GW200129}{\ensuremath{2.11\times10^{-24}}}%
		{GW200202}{\ensuremath{> \OTHREEOPA{}}}%
		{GW200208q}{--}%
		{GW200208am}{--}%
		{GW200209}{--}%
		{GW200210}{--}%
	}[\PackageError{OTHREEGWFAR}{Undefined option: #1}{}]
}%

\newcommand{\OTHREEGWPASTRO}[1]{%
	\IfEqCase{#1}{%
		{GW200112}{\ensuremath{>0.99}}%
		{GW200115}{\ensuremath{>0.99}}%
		{GW200128}{--}%
		{GW200129}{\ensuremath{>0.99}}%
		{GW200202}{--}%
		{GW200208q}{--}%
		{GW200208am}{--}%
		{GW200209}{--}%
		{GW200210}{--}%
	}[\PackageError{OTHREEGWPASTRO}{Undefined option: #1}{}]
}%

\newcommand{\OTHREEGWMCHIRP}[1]{%
	\IfEqCase{#1}{%
		{GW200112}{\ensuremath{35.37~M_{\odot}}}%
		{GW200115}{\ensuremath{2.57~M_{\odot}}}%
		{GW200128}{--}%
		{GW200129}{\ensuremath{32.74~M_{\odot}}}%
		{GW200202}{--}%
		{GW200208q}{--}%
		{GW200208am}{--}%
		{GW200209}{--}%
		{GW200210}{--}%
	}[\PackageError{OTHREEGWMCHIRP}{Undefined option: #1}{}]
}

\newcommand{\OTHREERETRACTIONS}{23}
\newcommand{\OTHREEGSTLALRETRACTIONS}{15}

\newcommand{\RETRACTIONFAR}{\ensuremath{1.67~\mathrm{per}~\mathrm{year}}}
\newcommand{\RETRACTIONSNR}{\ensuremath{14.5}}
\newcommand{\MDCRETRACTIONFARTHRESH}{one per year}

\newcommand{\BANKMASSLOW}{\ensuremath{1.0~M_{\odot}}}
\newcommand{\BANKMASSHIGH}{\ensuremath{200~M_{\odot}}}

\newcommand{\BHMASSLOW}{\ensuremath{3.0~M_{\odot}}}
\newcommand{\NSMASSLOW}{\ensuremath{1.0~M_{\odot}}}
\newcommand{\NSMASSHIGH}{$3.0 M_{\odot}$}
\newcommand{\TOTALMASSHIGH}{\ensuremath{400.0~M_{\odot}}}
\newcommand{\MASSRATIOHIGH}{\ensuremath{20}}

\newcommand{\NSSPIN}{\ensuremath{\lvert 0.05\rvert}}
\newcommand{\BHSPIN}{\ensuremath{\lvert 0.99\rvert}}
\newcommand{\CHIPBOUND}{\ensuremath{1\times10^{-3}}}

\newcommand{\MCHIRPBOUNDARY}{\ensuremath{1.73~M_{\odot}}}
\newcommand{\LOWMCHIRPWAVEFORM}{\texttt{TaylorF2}}
\newcommand{\HIGHMCHIRPWAVEFORM}{\texttt{SEOBNRv4}}

\section{Introduction}
\label{sec:intro}

Since the \ac{O1} of the \ac{LVK} Collaboration, \GSTLAL{}, a matched filtering based gravitational wave search pipeline~\cite{Messick:2016aqy}, has participated in the discovery of groundbreaking gravitational wave events. 
\GSTLAL{} was among the search pipelines that made the first direct detection of gravitational waves  from a merging \ac{BBH}, known as GW150914~\cite{LIGOScientific:2016aoc}. 
In the \ac{O2}, \GSTLAL{} was the first pipeline to observe the \ac{BNS} merger known as GW170817, whose discovery kickstarted the field of multi-messenger astronomy~\cite{GCN21505,LIGOScientific:2017vwq}.
In the \ac{O3}, \GSTLAL{} detected $\mathcal{O}(10)$s of gravitational wave signals including the first ever \ac{NSBH} mergers~\cite{LIGOScientific:2021qlt} and the very heavy \ac{BBH} merger, GW190521, which resulted in a remnant object in the \ac{IMBH} mass region~\cite{LIGOScientific:2020iuh}. 

The \GSTLAL{} pipeline can be operated in one of two configurations: a low-latency or ``online" mode and an ``offline" mode. 
The online configuration of the \GSTLAL{} analysis proceeds in near real time as strain data becomes available from the interferometers (currently, \ac{H1}, \ac{L1} and \ac{V1}). 
The online analysis enables the prompt detection of gravitational wave events, allowing for rapid communication to the external community for electromagnetic follow-up. 
In order to provide the best opportunities for multi-messenger astronomy, it is imperative that the low-latency analyses perform optimally.
This includes reliable signal recovery, accuracy of source property estimation, and the ability of the search to keep up with real-time data and provide results as quickly as possible.

In contrast, the offline analysis proceeds on long timescales relative to the low-latency distribution of strain data. 
The offline analysis can benefit from a fuller understanding of the detector noise and the ability to re-rank the significance of candidates against the full asynchronous background estimate collected over the entire run duration.
Since the likelihood ratios and \acp{FAR} of the candidates are re-computed relative to the full background, it is also possible to make adjustments to the signal model and mass model compared to what is used in the online analysis~\cite{Tsukada:2023edh}.
All of these factors can contribute to higher sensitivity, as quantified by the sensitive volume-time \VT{}, in the offline analysis. 

In this paper, we will focus on the online configuration and aim to characterize the \GSTLAL{} pipeline's performance toward the \ac{O4}.
In Section \ref{sec:software} we will describe the current configuration of the \GSTLAL{} online analysis. 
Additionally, we describe the \ac{GWLTS} software package as a tool for monitoring the performance of gravitational wave search pipelines in low-latency.
Then, in Section \ref{sec:mdc} we demonstrate the performance of the pipeline by presenting results from a \ac{MDC}.
We will conclude in Section \ref{sec:conclusion} with a description of on-going development towards \ac{O4}.

\section{Software description}
\label{sec:software}

\subsection{\GSTLAL{}}
\label{sec:gstlal}

\begin{figure}
\includegraphics[width=\columnwidth]{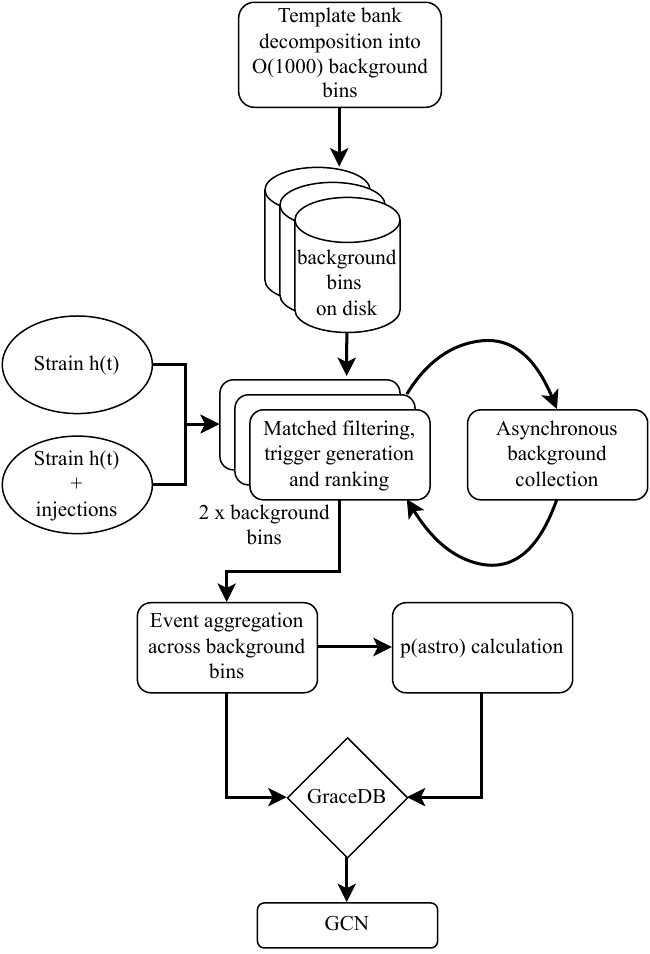}
\caption{\label{fig:online-diagram}
The low-latency \GSTLAL{} inspiral analysis workflow. 
The full template bank must first be decomposed into $\mathcal{O}(1000)$ independent background bins via the LLOID method of singular value decomposition and time slicing~\cite{Cannon:2011vi}.
The strain data is transferred from the interferometer sites at LIGO Livingston, LIGO Hanford, and Virgo to the computing clusters where it will be read from disk by the \GSTLAL{} pipeline.
Filtering, trigger generation, and candidate ranking proceeds in parallel for each background bin independently. 
These filtering and ranking jobs are duplicated to process strain channels which include simulated signals injected into the data.
Background statistics are collected independently in each background bin and asynchronously marginalized over the full parameter space in order to inform the \ac{FAR} estimation.
Candidate events are aggregated in time across all background bins, using the maximum \ac{SNR} or minimum \ac{FAR} as a metric for determining which candidates will be uploaded to GraceDB. 
Finally, candidates passing the public alert \ac{FAR} threshold will be disseminated via \ac{GCN}.
}
\end{figure}

The low-latency \GSTLAL{} inspiral workflow consists of two broad stages: a set up stage where pre-computed data products are generated and stored on disk and a persistent analysis stage where strain data is filtered in near real time and candidate events are identified. 
We will give a brief description of the current workflow and configuration choices to be used in operating the \GSTLAL{} analysis during \ac{O4}. 
A diagram of the low-latency workflow is shown in~\figref{fig:online-diagram}. 
For a more detailed description of the \GSTLAL{} analysis methods as of the end of \ac{O1} and \ac{O2} see~\cite{Messick:2016aqy} and~\cite{Sachdev:2019vvd}, respectively. 
The \GSTLAL{} software package is described in~\cite{Cannon:2020qnf}. 

Before the \GSTLAL{} analysis is launched, the template bank is first split into two halves in a process referred to as ``checkerboarding". 
Each checkerboarded bank is constructed by taking alternating neighboring templates from the full bank. 
The checkerboarded banks are redundant as they cover the same parameter space while having unique individual templates. 
The effectualness of the checkerboarded banks is validated in~\cite{Sakon:2022ibh}.
With this configuration the overall analysis can be split across two independent computing sites which improves the robustness of the analysis to upstream failures. 
According to the \ac{LLOID} method, the checkerboarded template banks must then be split into independent bins of waveforms, hereafter referred to as background bins, as shown in~\figref{fig:online-diagram}. 
A full derivation and motivation of the \ac{LLOID} method, including the \ac{SVD} and template time-slicing, is given by~\cite{Cannon:2011vi}.
We first sort the template bank by the orthogonal \ac{PN} phase terms $\mu_1$ and $\mu_2$.
These are linear combinations of the \ac{PN} coefficients, $\psi^0$, $\psi^2$, and $\psi^3$, defined as follows~\cite{Morisaki:2020oqk}:
\begin{equation} \label{mu}
\begin{split}
\mu_1 = 0.974  \psi^0 + 0.209 \psi^2 + 0.0840 \psi^3 \\
\mu_2 = -0.221 \psi^0 + 0.823 \psi^2 + 0.524 \psi^3.
\end{split}
\end{equation}
Using these parameters to sort, we split the template bank into sub-banks each with \TEMPLATESPERSUBBANK{} templates.
Each background bin is then constructed by grouping \NUMSUBBANKSPERSVD{} sub-banks together. 
When computing the decomposition, we require a \SVDTOLERANCE{} match between the re-constructed template waveforms and the initial physical waveforms. 
This value is chosen by balancing the need for computational efficiency with the need for accurately reconstructed waveforms. 
For the checkerboarded \ac{O4} template bank in~\cite{Sakon:2022ibh}, this produces \NUMSVDBANKS{} background bins.  

As part of the background bank construction during the set up stage of the analysis, the \ac{SVD} waveforms are also whitened. 
For the initial whitening before filtering has begun, we use a reference \ac{PSD}. 
As the analysis stage proceeds we will re-whiten the \ac{SVD} waveforms on a weekly timescale using recent \acp{PSD} in order to account for any long term changes to the detector characteristic noise.
As the analysis runs, the \ac{PSD} is continuously tracked using a \ac{FFT} length of \PSDFFTLENGTH{}. 
Such a short length of \ac{FFT} in the whitening stage of the pipeline reduces latency at the cost of a less accurate \ac{PSD} measurement which could potentially bring a loss in sensitivity while filtering.

The low-latency analysis ingests strain data, as well as data quality and interferometer state information from frame files. 
Each frame includes \FRAMELENGTH{} of data. 
The frames are distributed from the detectors via Apache Kafka, an open source event streaming platform. 
After streaming from the detector sites, frames are stored in shared memory partitions, where they are accessed by the \GSTLAL{} analysis. 
The frames are then processed in buffers \BUFFERBLOCKSIZE{} at a time by each filtering job in the \GSTLAL{} pipeline as shown in~\figref{fig:online-diagram}.
In \ac{O4}, the \GSTLAL{} pipeline will also ingest a parallel stream of strain data including simulated \ac{CBC} signals injected into the data. 
These injections will be based on the inferred astrophysical distribution of sources based on the \ac{GWTC}~\cite{KAGRA:2021duu}.

It is known that the LIGO and Virgo data are not ``well-behaved" and include non-transient and non-Gaussian noise components known as glitches. 
These glitches can be mistaken for astrophysical signals, especially high mass \ac{BBH} templates which are short in duration within the LIGO-Virgo frequency band. 
To mitigate the negative effects of non-Gaussian data, the \GSTLAL{} pipeline gates particularly glitchy whitened $h(t)$ strain data using a threshold on the amplitude of the data in units of standard deviations. 
In gating the strain data, we must be careful to balance the desire to reduce false positives (i.e. mistaking a glitch for an astrophysical signal) with the desire to avoid false negatives (i.e. mistaking an astrophysical signal for a glitch). 
Since we know that signals from heavy mass \ac{CBC} systems (for example, \ac{IMBH} binaries) tend to resemble glitches, we want to be conservative with gating data while filtering templates in this region of the parameter space. 
For this reason, we choose a mass-dependent gate threshold calculated for each background bin as follows. 
We first compute a gate ratio defined as:
\begin{equation}
R_{\mathrm{gate}} = \frac{\sigma_{\mathrm{max}} - \sigma_{\mathrm{min}}}{\mathcal{M}_{c,\mathrm{max}} - \mathcal{M}_{c,\mathrm{min}}}, 
\end{equation}
where we choose the minimum and maximum gate thresholds, $\sigma_{\mathrm{min}}$ and $\sigma_{\mathrm{max}}$ as \HTGATETHRESHOLDMIN{} and \HTGATETHRESHOLDMAX{}, respectively. 
The chirp mass, \MCHIRP{}, is defined by the following combination of the binary component masses: 
\begin{equation}
\MCHIRP= \frac{(m_1 m_2)^{3/5}}{(m_1 + m_2)^{1/5}}.
\end{equation}
The \MCHIRP{} minimum and maximum are taken to be $\mathcal{M}_{c,\mathrm{min}} = \HTGATEMCHIRPMIN{}$ and $\mathcal{M}_{c,\mathrm{max}} = \HTGATEMCHIRPMAX{}$. 
Then, the $h(t)$ gate threshold is calculated for each background bin as:
\begin{equation}
\sigma_{thr} = R_{\mathrm{gate}} \times \Delta \MCHIRP + \sigma_{\mathrm{min}}.
\end{equation}
Here, $\Delta \MCHIRP$ is the difference between the maximum \MCHIRP in the given background bin and the $\mathcal{M}_{c,\mathrm{min}}$.
This produces gate thresholds ranging from \HTGATEMIN{} for the smallest mass bins to \HTGATEMAX{} for the largest mass bins.

In the \ac{O4} configuration, we choose a filtering stride of \FIRSTRIDE{}, meaning that the matched filter output is computed in stretches of \FIRSTRIDE{} at a time.
The small stride is chosen to reduce latency in the filtering stage.
Triggers are defined by peaks in the \ac{SNR} time-series output by the matched filtering which pass a threshold of \TRIGGERSNRTHRESHOLD{}.
The \GSTLAL{} analysis has allowed for single detector candidates since \ac{O2} and will continue to do so in \ac{O4}. 
However, when calculating the likelihood ratio of single detector candidates we apply a penalty to down rank their significance.
This is a tunable parameter and the value to be used in \ac{O4} will be discussed in more detail in Sec.~\ref{sec:retractions}. 
Coincident candidates include triggers from the same template in at least two detectors.
We require that the end times of coincident triggers be within \COINCTHRESHOLD{} of each other after accounting for the light travel time between detectors. 
Together, the coincidence threshold and the requirement that triggers across detectors ring up the same template provide a strong signal consistency test for candidate events.

The \GSTLAL{} pipeline uses the likelihood ratio as a ranking statistic to assign significance to gravitational wave candidates~\cite{Cannon:2012zt,Cannon:2015gha}. 
Recent improvements to the likelihood ratio computation towards \ac{O4} are given in~\cite{Tsukada:2023edh}. 
These include an upgraded analytic \ac{SNR}-$\xi^2$ signal model and a method for removing signal contamination from the background which is also described in~\cite{Joshi:2023ltf}.
The background noise in each detector is estimated by collecting ranking statistic data from single-detector triggers observed in coincident time. 
We exclude triggers from times when only one detector is operating since these triggers may be astrophysical signals. 
These background estimations are cumulative and ``snapshotted" to disk every \LRSNAPSHOT{} hours. 
The filtering jobs which process injection strain data do not collect their own background estimations. 
This is because the high rate of injected signals in the data would contaminate the background and corrupt the statistics used for the \ac{FAR} estimation. 
Instead, these injection filtering jobs use a copy of the background statistics collected by the corresponding non-injection filtering job which processes the same background bin.

While the pipeline is designed to run persistently, there is need to take the analysis down periodically. 
We remove each of our analyses for a short period of time on a weekly timescale with a staggered schedule so that at least one of the checkerboarded analyses is always observing. 
When an analysis is re-launched after this weekly downtime, we compress the background ranking statistic data by removing any values in the horizon distance history that differ fractionally from their neighbors by less than \LRCOMPRESSION{}. 
This compression reduces the file size and memory use of the pipeline, which would otherwise grow without bounds over the duration of the observing period. 

For \ac{FAR} estimation, the ranking statistic data is marginalized by adding counts from the \ac{SNR}-$\xi^2$ background distributions collected in each background bin. 
The histograms are marginalized over in a continuous loop, taking several hours to complete each iteration. 
The marginalization is cumulative in time so that as the run proceeds, we collect more and more background counts. 
To account for the two redundant checkerboarded analyses, we apply a \ac{FAR} trials factor of \FARTRIALSFACTOR{} to each trigger. 

Gravitational wave events passing a \ac{FAR} threshold of one per hour will be uploaded to the \ac{GRACEDB}~\cite{gracedb}.
Because the \GSTLAL{} pipeline filters the strain data in \NUMSVDBANKS{} independent background bins, it is not only possible but highly probable that there will be multiple triggers associated with each physical gravitational wave candidate.
The number of triggers per candidate could range from a few for quieter signals to several tens for louder signals.
In order to reduce API calls to \ac{GRACEDB} we aggregate these triggers in time across background bins by the maximum \ac{SNR} and only upload the current best candidate. 
In this aggregation stage, triggers from different background bins are grouped into candidates using a coincidence window defined by rounding $t_{\mathrm{end}} - dt$ down to the nearest half second and rounding $t_{\mathrm{end}} + dt$ up to the nearest half second. 
Here, $t_\mathrm{end}$ is the end time of the trigger and $dt = \UPLOADDT{}$ seconds.
The first trigger received by the aggregator for a given candidate is uploaded to \ac{GRACEDB} immediately. 
Any subsequent triggers for the same candidate which are found with higher \ac{SNR} are uploaded with a \UPLOADCADENCE{} second geometric wait time. 
That is, after the first upload, the second upload will not be made until \UPLOADCADENCE{} seconds later, the third upload until $\UPLOADCADENCE{}^2$ seconds later, and so on. 
The aggregation stage of the pipeline is illustrated in~\figref{fig:online-diagram}. 

Finally, the \GSTLAL{} pipeline calculates a probability of astrophysical origin, or \PASTRO{}, for each event uploaded to \ac{GRACEDB}. 
The \PASTRO{} is a measure of the event's significance, and as we also compute the probability that the event originates from each \ac{CBC} source class (\ac{BNS}, \ac{NSBH}, or \ac{BBH}) it gives an indication of the likelihood that an event will have an electromagnetic counterpart. 
Therefore, the \PASTRO{} is an important quantity to help astronomers determine when to follow up gravitational wave candidates. 
More information about the \GSTLAL{} pipeline's computation of \PASTRO{} can be found in~\cite{Ray:2023nhx}.

\subsection{GW Low-latency Test Suite}
\label{sec:gwlts}

The \ac{GWLTS} software is designed to provide consistency checks and real-time feedback on the reliability of science outputs of gravitational wave search pipelines. 
By using simulated signals injected in the strain data, we can compare the pipeline performance to what is expected. 

The Test Suite requires a source of truth for the signals that are present in the data. 
For this, we rely on an injection set on disk which defines all of the injections, including all intrinsic and extrinsic parameters, and the \ac{GPS} times at which they appear in the strain data. 
Using a live estimate of the \ac{PSD} and the injected signal's sky location we can compute the expected \ac{SNR}. 
For information about recovered events, \ac{GRACEDB} is taken as the source of truth. 
The \IGWNALERT{} software package is a messaging system built on Apache Kafka which sends notifications of \ac{GRACEDB} state changes to subscribed users.
The Test Suite subscribes to notifications from \IGWNALERT{} which are sent for any new or updated event on \ac{GRACEDB}. 
The injections are then matched with recovered alerts in low-latency by finding coincidences within a small $\Delta t$ which we take to be \TESTSUITECOINCWINDOW{} seconds. 
This time window was chosen to be very small compared to a typical injection rate to avoid erroneous coincidences. 

Once an injected signal is matched with a recovered event, the information is passed to an arbitrary number of independent jobs via Apache Kafka. 
The jobs compute metrics associated with the injection recovery such as the \VT{}, accuracy of source classification and sky localization, and accuracy of point estimates of the source intrinsic parameters. 
The \ac{GWLTS} capabilities are described in further detail in Section \ref{sec:mdc}. 

All of the metrics computed by the \ac{GWLTS} are stored with InfluxDB, which is an open source platform for storing and querying time series data. 
We use the data visualization tool Grafana to display the data in real time in online dashboards. 
With this infrastructure, we are able to track changes in the performance of the analysis on the timescale of seconds. 
Additionally, from the Influx database we are able to keep an archival record of the performance metrics.

\section{Mock data challenge results}
\label{sec:mdc}

To demonstrate the performance of the \GSTLAL{} analysis and our readiness for \ac{O4}, we participated in a \ac{MDC} consisting of a forty day stretch of \ac{H1}\ac{L1}\ac{V1} \ac{O3} strain data taken from \MDCSTART{} to \MDCEND{} UTC and replayed so as to be analyzed in a low-latency configuration. 
The \ac{MDC} also provided a set of identical strain channels with injected \ac{BNS}, \ac{NSBH}, and \ac{BBH} signals. 
Details of the injection distributions used in the \ac{MDC} can be found in~\cite{mdc_analytics}. 
Injected signals were placed in the strain data at a rate of one per \INJECTIONSPACING{} seconds, leading to a total of \TOTALINJECTIONS{} total injections throughout the \ac{MDC} duration.

In this section we seek to quantify the performance of the \GSTLAL{} pipeline in its latest configuration.
We will first show the recovery of known gravitational wave events in the \ac{MDC} data, as well as highlight any potential retraction level events.
We will then detail the results of the \ac{MDC} injection campaign.
Finally we present the stability and performance of the pipeline in terms of its uptime and latency.
 
\subsection{Gravitational wave events}
\label{sec:gw-events}

There are $9$ gravitational wave events in the duration of the \ac{MDC} replay data which were previously published as significant candidates in \ac{GWTC}~\cite{LIGOScientific:2021djp}. 
These are described throughout the remainder of this section and summarized in Table~\ref{tab:gw-events}, comparing the \GSTLAL{} pipeline's recovery of the signal in \ac{O3} to that in the \ac{MDC}.
We recover all of the $9$ candidates at the \HIGHFARTHRESH{} \ac{FAR} threshold for uploading to \ac{GRACEDB}.
Of these, three were found with high significance by \GSTLAL{} in the \ac{O3} online analysis. 
Two were found with marginal or sub-threshold significance online but with high significance offline. 
Four candidates were only found by \GSTLAL{} in the offline analysis.
The recovery of all previously published candidates shows that the pipeline is performing with at least the same capability as in \ac{O3}.

\begin{event_table}
\begin{table*}
{\small
\noindent\begin{tabularx}{\textwidth}{l@{\extracolsep{\fill}} cccc cccc}
 & \multicolumn{4}{c}{\textbf{\ac{O3} Online}} & \multicolumn{4}{c}{\textbf{\ac{MDC}}} \\
\cline{2-5} \cline{6-9} \\
\textbf{Name} & \textbf{Inst} & \textbf{\ac{SNR}} & \textbf{FAR} & $\boldsymbol{\PASTRO{}}$ & \textbf{Inst} & \textbf{\ac{SNR}} & \textbf{FAR} & $\boldsymbol{\PASTRO{}}$ \\
& & & ($\mathrm{yrs}^{-1}$) & & & & ($\mathrm{yrs}^{-1}$) & \\
\hline
\makebox[0pt][l]{\fboxsep0pt\colorbox{lightgray}{\mystrut\hspace*{1.0\linewidth}}} GW$200112\_155838$ & \OTHREEGWIFOS{GW200112} & \OTHREEGWSNR{GW200112} & \OTHREEGWFAR{GW200112} & \OTHREEGWPASTRO{GW200112} & \MDCGWIFOS{GW200112} & \MDCGWSNR{GW200112} & \MDCGWFAR{GW200112} & \MDCGWPASTRO{GW200112} \\
GW$200115\_042309$ & \OTHREEGWIFOS{GW200115} & \OTHREEGWSNR{GW200115} & \OTHREEGWFAR{GW200115} & \OTHREEGWPASTRO{GW200115} & \MDCGWIFOS{GW200115} & \MDCGWSNR{GW200115} & \MDCGWFAR{GW200115} & \MDCGWPASTRO{GW200115} \\
\makebox[0pt][l]{\fboxsep0pt\colorbox{lightgray}{\mystrut\hspace*{1.0\linewidth}}} GW$200128\_022011$ & \OTHREEGWIFOS{GW200128} & \OTHREEGWSNR{GW200128} & \OTHREEGWFAR{GW200128} & \OTHREEGWPASTRO{GW200128} & \MDCGWIFOS{GW200128} & \MDCGWSNR{GW200128} & \MDCGWFAR{GW200128} & \MDCGWPASTRO{GW200128} \\
GW$200129\_065458$ & \OTHREEGWIFOS{GW200129} & \OTHREEGWSNR{GW200129} & \OTHREEGWFAR{GW200129} & \OTHREEGWPASTRO{GW200129} & \MDCGWIFOS{GW200129} & \MDCGWSNR{GW200129} & \MDCGWFAR{GW200129} & \MDCGWPASTRO{GW200129} \\
\makebox[0pt][l]{\fboxsep0pt\colorbox{lightgray}{\mystrut\hspace*{1.0\linewidth}}} GW$200202\_154313$ & \OTHREEGWIFOS{GW200202} & \OTHREEGWSNR{GW200202} & \OTHREEGWFAR{GW200202} & \OTHREEGWPASTRO{GW200202} & \MDCGWIFOS{GW200202} & \MDCGWSNR{GW200202} & \MDCGWFAR{GW200202} & \MDCGWPASTRO{GW200202} \\
GW$200208\_130117$ & \OTHREEGWIFOS{GW200208q} & \OTHREEGWSNR{GW200208q} & \OTHREEGWFAR{GW200208q} & \OTHREEGWPASTRO{GW200208q} & \MDCGWIFOS{GW200208q} & \MDCGWSNR{GW200208q} & \MDCGWFAR{GW200208q} & \MDCGWPASTRO{GW200208q} \\
\makebox[0pt][l]{\fboxsep0pt\colorbox{lightgray}{\mystrut\hspace*{1.0\linewidth}}} GW$200208\_222617$ & \OTHREEGWIFOS{GW200208am} & \OTHREEGWSNR{GW200208am} & \OTHREEGWFAR{GW200208am} & \OTHREEGWPASTRO{GW200208am} & \MDCGWIFOS{GW200208am} & \MDCGWSNR{GW200208am} & \MDCGWFAR{GW200208am} & \MDCGWPASTRO{GW200208am} \\
GW$200209\_085452$ & \OTHREEGWIFOS{GW200209} & \OTHREEGWSNR{GW200209} & \OTHREEGWFAR{GW200209} & \OTHREEGWPASTRO{GW200209} & \MDCGWIFOS{GW200209} & \MDCGWSNR{GW200209} & \MDCGWFAR{GW200209} & \MDCGWPASTRO{GW200209} \\
\makebox[0pt][l]{\fboxsep0pt\colorbox{lightgray}{\mystrut\hspace*{1.0\linewidth}}} GW$200210\_092254$ & \OTHREEGWIFOS{GW200210} & \OTHREEGWSNR{GW200210} & \OTHREEGWFAR{GW200210} & \OTHREEGWPASTRO{GW200210} & \MDCGWIFOS{GW200210} & \MDCGWSNR{GW200210} & \MDCGWFAR{GW200210} & \MDCGWPASTRO{GW200210} \\
\hline
\end{tabularx}
}
\caption{
Gravitational wave candidates from \MDCSTART{} to \MDCEND{} UTC as recovered by the \GSTLAL{} pipeline during the \ac{O3} online analysis and during the \ac{MDC}. 
The instruments provided are those which participated in the event, that is, contributed a trigger with \ac{SNR} $>\TRIGGERSNRTHRESHOLD{}$. 
Here, the \ac{SNR} is the recovered network \ac{SNR}, \ac{FAR} is the false alarm rate in inverse years, and \PASTRO{} is the probability of astrophysical origin. 
The two low-significance candidates identified with \ac{FAR} above the public alert threshold in the \ac{O3} online analysis are indicated with \ac{FAR} $> \OTHREEOPA{}$ per year. 
We note that this \ac{FAR} threshold is after a trials factor corresponding to the number of operating pipelines has been applied. 
The last four candidates in the table were not recovered by \GSTLAL{} in the \ac{O3} online analysis. 
}
\label{tab:gw-events}
\end{table*}
\end{event_table}

\begin{figure}
\includegraphics[scale=0.5]{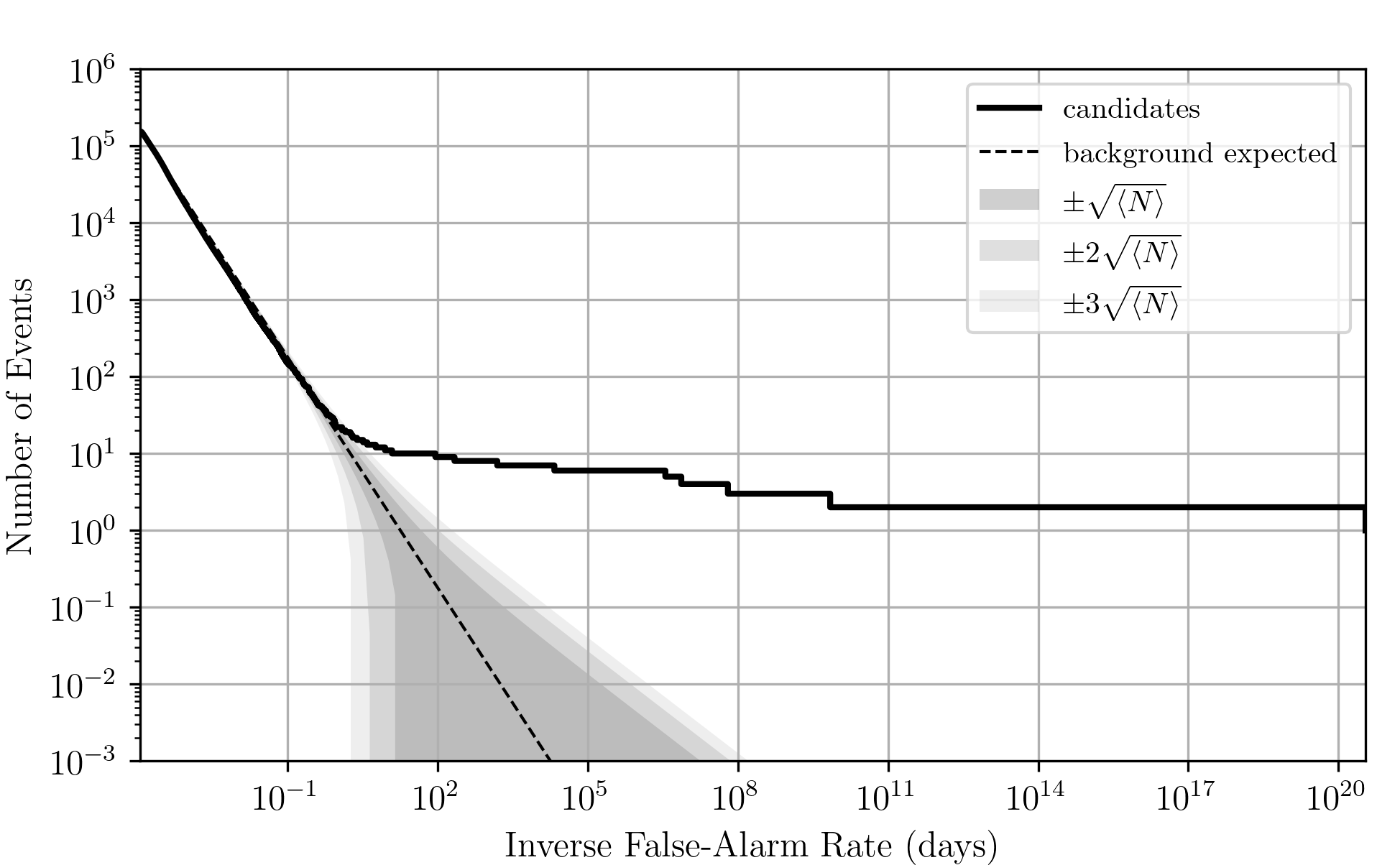}
\caption{\label{fig:money-plot}
Count of observed candidates vs \ac{IFAR} in days.
The dashed line is an estimation of the expected number of background counts, assuming a \ac{FAR} threshold of $1\times10^4$ per day.
The one, two, and three $\sigma$ error regions are indicated by the shaded regions.}
\end{figure}

~\figref{fig:money-plot} shows the count of observed candidates versus \ac{IFAR}. 
The expected background counts are calculated using an estimated livetime which is equal to the wall clock time from the first to the last candidate. 
Additionally, we apply a trials factor of \FARTRIALSFACTOR{} to the \acp{FAR} since we only include candidates from one of the checkerboarded analyses. 
~\figref{fig:money-plot} shows the $9$ known gravitational wave events recovered in the \ac{MDC}. 
The candidate \ac{IFAR} statistics agree with the expected counts from noise at lower \ac{IFAR} and diverge due to the presence of signals at higher \ac{IFAR}.

$\mathbf{\mathrm{\textbf{GW}}200112\_155838}.$ 
This event was a \ac{BBH} candidate observed in LIGO Livingston data as a single detector candidate with chirp mass \OTHREEGWMCHIRP{GW200112} in \ac{O3} and \MDCGWMCHIRP{GW200112} in the \ac{MDC}. 
In the \ac{MDC}, we recover the event with a comparable \ac{SNR} as that observed in \ac{O3}, however in the \ac{MDC} the \ac{FAR} is significantly lower.

$\mathbf{\mathrm{\textbf{GW}}200115\_042309}$ was the first confident \ac{NSBH} detection found in \ac{O3}. 
The event was found as a coincident trigger in LIGO Hanford and LIGO Livingston data.
The \ac{SNR} recovery, \ac{FAR} estimation, and chirp mass estimation are all equivalent in the \ac{MDC} to what was observed in \ac{O3}.

$\mathbf{\mathrm{\textbf{GW}}200129\_065458}$ was a \ac{BBH} and the loudest gravitational wave signal in the duration of the \ac{MDC} with \ac{O3} \ac{SNR} $= \OTHREEGWSNR{GW200129}$. 
The event was recovered well below the public alert threshold in both \ac{O3} and the \ac{MDC}. 


$\mathbf{\mathrm{\textbf{GW}}200128\_022011}$ and $\mathbf{\mathrm{\textbf{GW}}200202\_154313}$ are both \ac{BBH} candidates found by \GSTLAL{} in the \ac{O3} online analysis with low significance.
Both candidates were found with \ac{FAR} above the \ac{O3} public alert threshold of \OTHREEOPA{} per year, where a trials factor corresponding to the number of operating pipelines has been applied. 
Later, during the offline analysis they were recovered as significant candidates and included in \ac{GWTC}~\cite{LIGOScientific:2021djp}.
In the \ac{MDC} we recover both candidates with significantly lower \acp{FAR}, both well below the public alert threshold. 
Therefore, if similar events occur during \ac{O4}, we can expect to recover them as significant public alerts. 


$\mathbf{\mathrm{\textbf{GW}}200208\_130117}$ was not recovered by \GSTLAL{} in the \ac{O3} online analysis, however it was found in the offline analysis by \GSTLAL{} as a marginally significant candidate~\cite{LIGOScientific:2021djp}. 
As recovered in the \ac{MDC}, this event is a \ac{BBH} candidate with chirp mass \MDCGWMCHIRP{GW200208q} and a much lower \ac{FAR} than what was found in either the \ac{O3} online or offline analyses. 

$\mathbf{\mathrm{\textbf{GW}}200208\_222617}$ was only recovered by \GSTLAL{} as a sub-threshold candidate in the \ac{O3} offline analysis, and its inclusion as a significant candidate in \ac{GWTC} was due to its recovery by other \ac{CBC} pipelines~\cite{LIGOScientific:2021djp}. 
The \GSTLAL{} pipeline did not recover this event in \ac{O3} online. 
In the \ac{MDC} the event was recovered with low significance at \ac{SNR} $=\MDCGWSNR{GW200208am}$ and \ac{FAR} $=\MDCGWFAR{GW200208am}$ per year. 

$\mathbf{\mathrm{\textbf{GW}}200209\_085452}$ and $\mathbf{\mathrm{\textbf{GW}}200210\_092254}$  were not recovered by \GSTLAL{} in the \ac{O3} online analysis, however they were found in the offline analysis by \GSTLAL{} with high significance and marginal significance, respectively.
In the \ac{MDC}, $\mathrm{GW}200209\_085452$ was recovered as a high significance candidate with \ac{FAR} $=\MDCGWFAR{GW200209}$ per year. 
This event is a \ac{BBH} candidate with chirp mass $=\MDCGWMCHIRP{GW200209}$ found in LIGO Hanford and LIGO Livingston data.
$\mathrm{GW}200210\_092254$ was found as a sub-threshold candidate in the \ac{MDC} with \ac{FAR} $=\MDCGWFAR{GW200210}$ per year. 
If astrophysical, the event would be an \ac{NSBH} candidate with chirp mass $=\MDCGWMCHIRP{GW200210}$. 

The improved performance of the \GSTLAL{} pipeline in the \ac{MDC} as compared to the \ac{O3} online analysis can be attributed to a number of incremental improvements made to the likelihood ratio ranking statistic and background estimation. 
~\cite{Tsukada:2023edh} describes an improved signal model and~\cite{Joshi:2023ltf} introduces a new method for a time-dependent background wherein contamination is reduced by removing signals counts from the background $\ac{SNR}-\xi^2$ histograms. 
Each of these changes have introduced a small improvement to the \VT{} which, when combined, leads to a noticeable increase in sensitivity and corresponding number of detected events.

\subsection{Retractions}
\label{sec:retractions}

\begin{figure}
\includegraphics[scale=0.3]{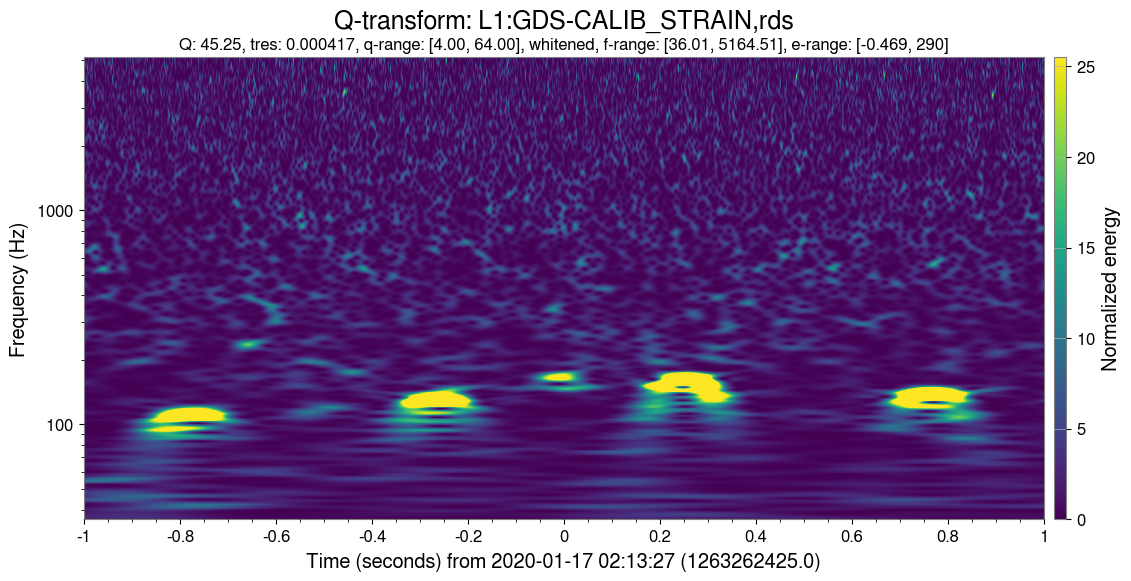}
\caption{\label{fig:retraction-q-scans}
Spectrogram of L1 $h(t)$ data for $\pm 1$ second around the time of the retraction level candidate recovered in the \ac{MDC}.
This candidate is expected to be terrestrial in origin due to the clear presence of glitches in this data.
}
\end{figure}

In \ac{O3}, there were \OTHREERETRACTIONS{} public gravitational wave candidates which were subsequently determined to be terrestrial in origin and thus retracted.
Of these, \GSTLAL{} contributed to \OTHREEGSTLALRETRACTIONS{}.
In \ac{O4}, we hope to significantly reduce the number of retractions produced by the \GSTLAL{} pipeline.
Four of the \OTHREERETRACTIONS{} retractions took place during the stretch of data covered by the \ac{MDC}. 
These are: S200106au, S200106av, S200108v, and S200116ah~\cite{GCN26641,GCN26665,GCN26785}.

\GSTLAL{} did not upload triggers for S200106au and S200106av during \ac{O3}. 
In the \ac{MDC}, these events would have occurred at a time before the pipeline had collected sufficient background to begin ranking candidates, and thus we did not upload triggers for these events.
The retractions S200108v and S200116ah were \GSTLAL{}-only candidates in \ac{O3}, both being found as \ac{L1} single detector candidates. 
Again, the time corresponding to S200108v would have been early enough in the \ac{MDC} cycle that the pipeline was not ranking or uploading triggers yet, so we cannot make any comparison to our performance in \ac{O3} for this retraction. 
Finally, we did not produce any trigger below the \HIGHFARTHRESH{} \ac{FAR} threshold for uploading to \ac{GRACEDB} corresponding to S200116ah in the \ac{MDC}, despite the analysis being fully burned in and operating in a stable state. 
This indicates an improvement in our rate of retractions for \ac{O4}, however there is not enough data within the \ac{MDC} to make a strong statement. 

In addition to the retracted candidates uploaded by \GSTLAL{} in \ac{O3}, for the purpose of the \ac{MDC} we define a ``retraction level candidate" as any gravitational wave candidate uploaded with a \ac{FAR} less than \MDCRETRACTIONFARTHRESH{} which is not in the list of previously published candidates discussed earlier in this section. 
Over the duration of the \ac{MDC}, we find one such retraction level candidate.
This was a single-detector candidate found in \ac{L1} with an \ac{SNR} of \RETRACTIONSNR{}.
The recovered \ac{FAR} was low enough to be counted as a significant candidate with \ac{FAR} $ = \RETRACTIONFAR{}$.
The \ac{L1} data around the event time shows the clear presence of scattering glitches, as shown in ~\figref{fig:retraction-q-scans}.
Further evidence of terrestrial origin for this candidate is that no coincident triggers were recovered in \ac{H1} or \ac{V1} despite both of these detectors operating normally at the time.
Candidates recovered in only a single detector are more susceptible to uncertainty since they lack the strong signal consistency test of coincidence in multiple detectors. 
For this reason, there has been a penalty applied to the ranking statistic of single detector candidates which down weights their significance. 
In the \ac{O3} offline analysis and in the \ac{MDC} we used a singles penalty of \SINGLESPENALTYMDCELEVEN{} in log likelihood ratio, however in order to reduce the number of similar retraction level events in \ac{O4}, we plan to increase the singles penalty to \SINGLESPENALTYOFOUR{}.
With this penalty applied, the retraction event in the \ac{MDC} would be down weighted and expected to be recovered with a \ac{FAR} greater than two per year.

\subsection{Recovered injections}

\begin{figure}
    \includegraphics[scale=0.35]{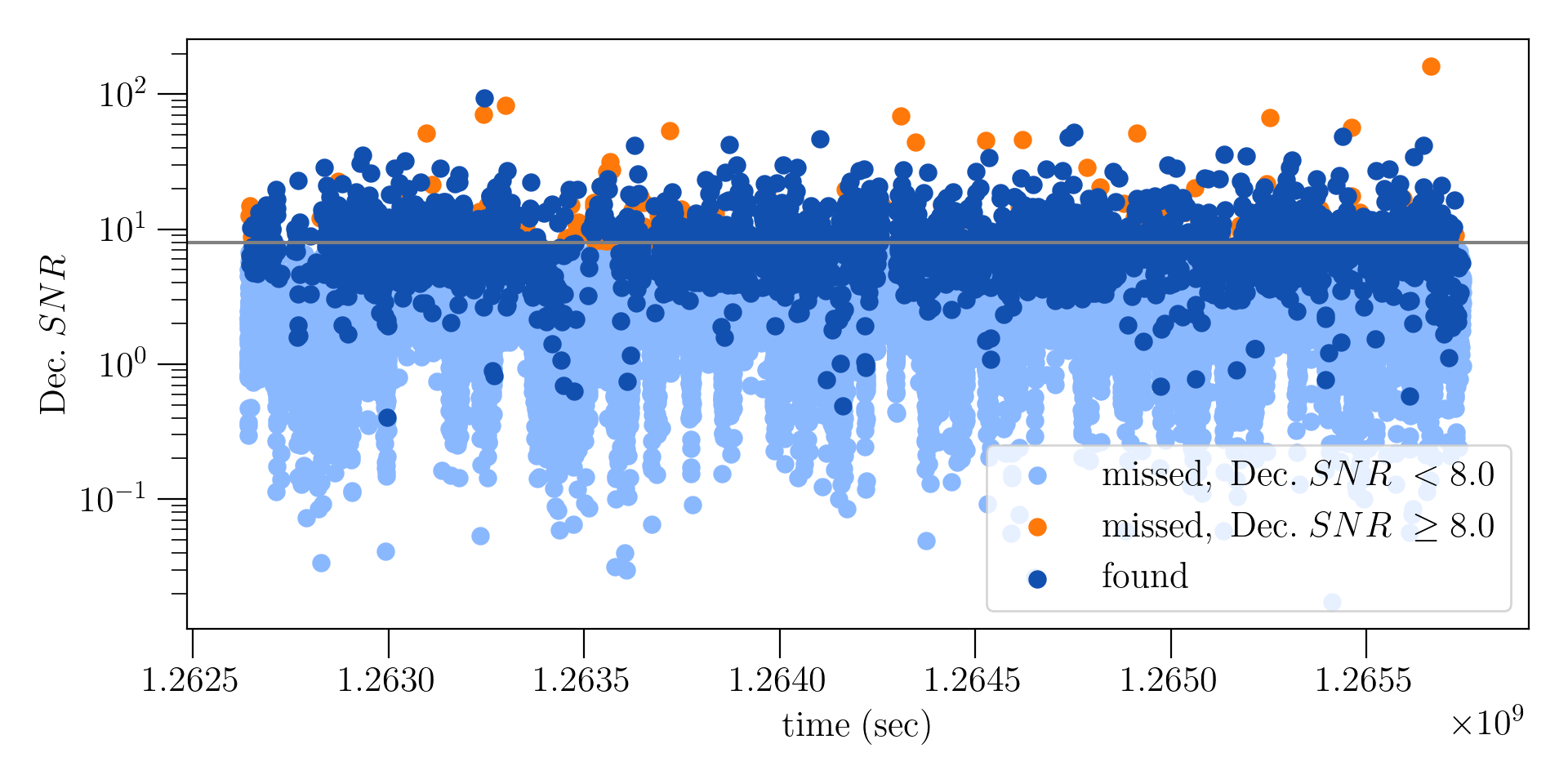}
    \caption{\label{fig:dec-snr}
Time-series of injected decisive \ac{SNR} for injections with component masses and spins within the \ac{O4} template bank.
Dark blue markers indicate injections that were recovered below a \LOWFARTHRESH{} \ac{FAR} threshold. 
Orange and light blue markers indicate injections not recovered below this \ac{FAR} threshold, where orange points are injections with decisive \ac{SNR} $>\DECISIVESNRTHRESH{}$. 
Times on the horizontal axis are GPS times shifted to the original \ac{O3} epoch. 
}
\end{figure}

There were \TOTALINJECTIONS{} simulated signals injected into the five week duration of the \ac{MDC} strain data. 
Of these, many had component masses and spins outside the region of parameter space covered by our template bank.
In addition to the injection parameters, the expected recovery of each injection is dependent on the set of interferometers producing science quality data at the time of the injection.
During times when no interferometers are operating we of course do not expect to recover any injections.
We define the decisive \ac{SNR} as the \ac{SNR} in the second most sensitive interferometer during times when multiple interferometers were observing, and the only available SNR otherwise.
The decisive \ac{SNR} is a more informative measure of the loudness of an injection than the network \ac{SNR} since it wraps in information about the set of operating interferometers. 
While all injections have network \ac{SNR} $\geq 4.0$, we find that many injections have decisive \ac{SNR} $< 4.0$. 

~\figref{fig:dec-snr} shows the time-series of decisive \ac{SNR} for all injections throughout the \ac{MDC}.
For the purpose of this paper, we focus on injections whose parameters fall inside our bank.
That is, injections with component masses between \BANKMASSLOW{} and \BANKMASSHIGH{}, with total masses, $m_1 + m_2 <~\TOTALMASSHIGH{}$ and mass ratios, $q = m_1 / m_2 <~\MASSRATIOHIGH{}$.
For objects with mass $< \BHMASSLOW{}$ the template bank restricts spins perpendicular to the orbital plane to be $< \NSSPIN{}$, and for objects with mass $> \BHMASSLOW{}$ allows spins up to \BHSPIN{}.
We use the effective precession spin, \CHIP{}, defined in~\cite{Schmidt:2014iyl} as:
\begin{equation}
\chi_p = \frac{\mathrm{max}(a_1 \cdot s_1, a_2 \cdot s_2)}{a_1 \cdot m_1^2}
\end{equation}
to quantify the in-plane spin of injections.
Here, $a_1 = 2 + 3 / 2q$, $a_2 = 2 + 3q / 2$, and $s_i = \sqrt{s_{i,x}^2 + s_{i,y}^2}$. 
And $q$ is the mass ratio, taking $m_1 \ge m_2$.
Since the template bank does not include any in-plane spins, we focus on injections with $\chi_p < \CHIPBOUND{}$.
However, we find that all injections with one component mass $< \BHMASSLOW{}$ have spins outside the range of the bank, therefore we relax the spin conditions on these components.
The mass and spin restrictions that we use are summarized in Table~\ref{tab:in-bank}.
Finally, to account for the fact that not all interferometers were providing science quality data at all times, we highlight injections with an estimated decisive \ac{SNR} $ \ge \DECISIVESNRTHRESH{}$.
These cuts leave a total of \ALLINBANKABOVEDECSNRTHRESH{} injections during the five week \ac{MDC}. 
Of these, there are \BBHINBANKABOVEDECSNRTHRESH{} \ac{BBH}, \BNSINBANKABOVEDECSNRTHRESH{} \ac{BNS}, and \NSBHINBANKABOVEDECSNRTHRESH{} \ac{NSBH} injections.

\begin{figure}[h]
\includegraphics[scale=0.5]{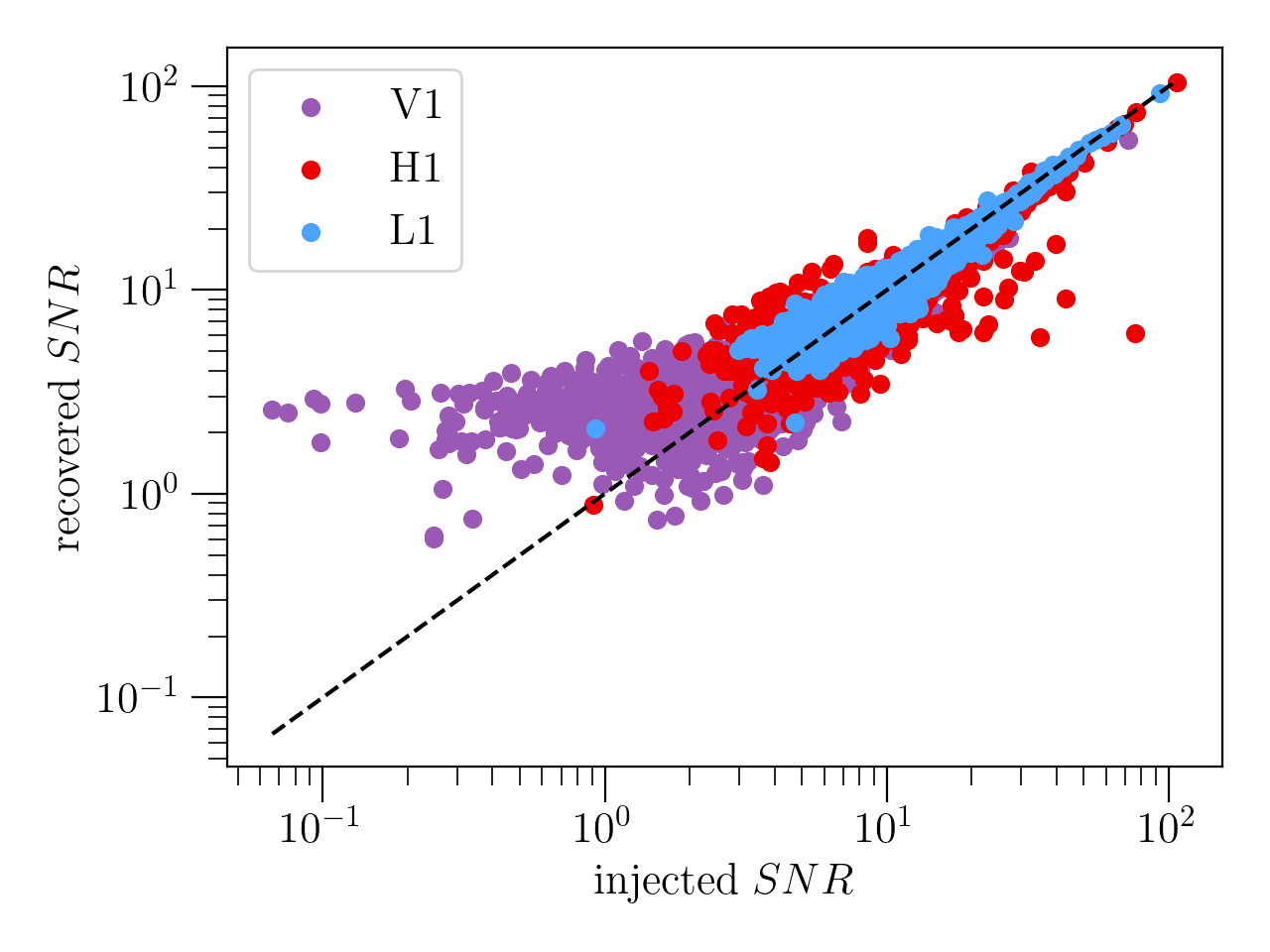}
\caption{\label{fig:snr-acc}
Recovered and estimated injected \ac{SNR} in each interferometer: \ac{H1}, \ac{L1}, and \ac{V1}.
}
\end{figure}

The injected \acp{SNR} are not known in advance of the \ac{MDC}, but we estimate them using \ac{GWLTS}.
We calculate the injected strain time series using the injection end time, sky position, and other source intrinsic parameters assuming an \texttt{IMRPhenomPv2\_NRTidalv2} waveform~\cite{Dietrich:2019kaq}. 
We use a running estimate of the detector \acp{PSD} and estimate the \ac{SNR} with a lower (upper) frequency cut off of \INJSNRFLOW{} (\INJSNRFHI{}) Hz. 
~\figref{fig:snr-acc} shows the recovered and estimated injected \ac{SNR} for each detector. 
If the template bank did not have a sufficient minimal match, we may expect to see systematically lower recovered \acp{SNR} than the expected values. 
However, we find that the recovered \ac{SNR} generally aligns with the expected \ac{SNR}.

\begin{event_table}
\begin{table}
{\small
\noindent\begin{tabularx}{\columnwidth}{l@{\extracolsep{\fill}}rrrrrrrr}
$\boldsymbol{m_1}$ & $\boldsymbol{m_2}$ & $\boldsymbol{M} $& $\boldsymbol{q}$ & $\boldsymbol{s_{1,z}}$ &  $\boldsymbol{s_{2,z}}$ & $\boldsymbol{\CHIP{}}$ \\
\hline
\makebox[0pt][l]{\fboxsep0pt\colorbox{lightgray}{\mystrut\hspace*{1.0\linewidth}}} $1,3$ & $1,3$ & $<6$ & $< 0.33$ & -- & -- & -- \\
$ 3,200$ & $1,3$ & $<203$ & $< 20$ & --  & -- & -- \\
\makebox[0pt][l]{\fboxsep0pt\colorbox{lightgray}{\mystrut\hspace*{1.0\linewidth}}} $ 3,200$ & $3,200$ & $<400$ & $<20$ & $<0.99$  & $<0.99$ & $< 0.001$ \\
\hline
\end{tabularx}
}
\caption{Restrictions on parameters of injections according to the \ac{O4} template bank boundaries. 
The ``--" indicates that we make no restrictions on the given parameter for injections (even though there may be restrictions on these parameters imposed by the template bank).
}
\label{tab:in-bank}
\end{table}
\end{event_table}

An injection is considered ``found" if it is recovered by the pipeline with a \ac{FAR} passing some pre-determined threshold, and ``missed" otherwise. 
We will quote most results in the following sections with respect to a \LOWFARTHRESH{} \ac{FAR} threshold.
At the time of writing, this is the threshold expected to be used in \ac{O4} for sending public alerts~\cite{user_guide}.
However, the \acp{FAR} of \ac{CBC} signals will be subject to a trials factor corresponding to the number of operating pipelines, so the effective alert threshold will be lower. 
We define the injection recovery efficiency as,
\begin{equation}
\label{eq:eff}
\frac{\mathrm{found\; injections}}{(\mathrm{found\; injections} + \mathrm{missed\; injections})}.
\end{equation}

At the \LOWFARTHRESH{} \ac{FAR} threshold the efficiency was $ = \ALLINBANKEFFICIENCY{TWOPERDAY}$ for all injections in the template bank.
The recovered injection efficiencies for each source class are shown in Table~\ref{tab:efficiency} at four typical \ac{FAR} thresholds.
As is expected, the efficiencies are better at more conservative \ac{FAR} thresholds.
The analysis has the highest efficiency for injections consistent with \ac{BNS} sources, and the lowest efficiency for \ac{NSBH} sources. 

We would expect the pipeline to recover all injections above some decisive \ac{SNR} or network \ac{SNR} threshold. 
However, ~\figref{fig:dec-snr} shows that there are several very high \ac{SNR} missed injections throughout the duration of the \ac{MDC}. 
We find that most of the missed injections with decisive \ac{SNR} $> 20.0$ are high mass \ac{BBH} injections and a few are high mass ratio \ac{NSBH} injections.
This results in a decrease of the \ac{BBH} recovery efficiency as the injections increase in \ac{SNR}, which is contradictory to our expectations. 
These injections are missed due to falling outside of the \ac{SNR}-$\xi^2$ signal region used in the likelihood ratio calculation.
The signal region is an analytic model which depends on the allowed mismatch\footnote{The ``mismatch" can also mean the fractional loss in \ac{SNR} due to differences between the template parameters and the true waveform. However, here we refer to the mismatch as defined in~\cite{Tsukada:2023edh} which is an unnormalized quantity, therefore retaining a dependence on the \ac{SNR}.} between recovered \ac{SNR} time-series and the template waveform as part of the autocorrelation $\xi^2$ test.
If the allowed mismatch range is too strict, it will result in a narrow signal model which can exclude real signals. 
This effect is exaggerated at high \ac{SNR} where we expect larger mismatches due to the discreteness of the template bank as well as waveform systematics.
In the \ac{MDC}, we used a mismatch range of \XISQMISMATCHRANGE{}. 
The optimal mismatch range in the signal model is an open area of study. 
See~\cite{Tsukada:2023edh} for a more detailed discussion. 

\begin{event_table}
\begin{table}
{\small
\noindent\begin{tabularx}{\columnwidth}{l@{\extracolsep{\fill}}c c c c}
 \textbf{\ac{FAR}} & \textbf{\ac{BNS}} & \textbf{\ac{NSBH}} & \textbf{\ac{BBH}} & \textbf{ALL} \\
\hline
\makebox[0pt][l]{\fboxsep0pt\colorbox{lightgray}{\mystrut\hspace*{1.0\linewidth}}} \ONEPERHOUR{} & \BNSINBANKEFFICIENCY{ONEPERHOUR} &\NSBHINBANKEFFICIENCY{ONEPERHOUR} & \BBHINBANKEFFICIENCY{ONEPERHOUR} & \ALLINBANKEFFICIENCY{ONEPERHOUR} \\
\TWOPERDAY{} & \BNSINBANKEFFICIENCY{TWOPERDAY} & \NSBHINBANKEFFICIENCY{TWOPERDAY} & \BBHINBANKEFFICIENCY{TWOPERDAY} & \ALLINBANKEFFICIENCY{TWOPERDAY} \\
\makebox[0pt][l]{\fboxsep0pt\colorbox{lightgray}{\mystrut\hspace*{1.0\linewidth}}} \ONEPERMONTH{} & \BNSINBANKEFFICIENCY{ONEPERMONTH} & \NSBHINBANKEFFICIENCY{ONEPERMONTH} & \BBHINBANKEFFICIENCY{ONEPERMONTH} & \ALLINBANKEFFICIENCY{ONEPERMONTH} \\ 
\TWOPERYEAR{} & \BNSINBANKEFFICIENCY{TWOPERYEAR} & \NSBHINBANKEFFICIENCY{TWOPERYEAR} & \BBHINBANKEFFICIENCY{TWOPERYEAR} & \ALLINBANKEFFICIENCY{TWOPERYEAR} \\
\hline
\end{tabularx}
}
\caption{
Injection efficiencies as defined in Eq.~\ref{eq:eff} computed using four \ac{FAR} thresholds to count ``found" injections: one per hour (the \ac{GRACEDB} upload threshold), two per day (the public alert threshold), one per month, and two per year. 
Source categories are defined in Table~\ref{tab:in-bank} and ``ALL" combines injections from the three source categories.
}
\label{tab:efficiency}
\end{table}
\end{event_table}

\subsubsection{Injection parameter recovery}

In this section we will quantify the accuracy of point estimates of the source intrinsic parameters made by the \GSTLAL{} pipeline.
These estimates simply come from the template parameters of the trigger which rang up the maximum \ac{SNR} across background bins.
An understanding of the parameter accuracy obtained by search pipelines can be useful to full parameter estimation efforts. 
For example, the Bayesian inference library Bilby~\cite{Smith:2019ucc,Ashton:2018jfp} relies on the choice of prior probability distributions for intrinsic parameters. 
When parameters are well determined by the searches, Bilby can use narrow distributions around those values, otherwise more broad prior distributions must be used.
We present parameter accuracy results for the chirp mass \MCHIRP{}, effective inspiral spin \CHIEFF{}, mass ratio $q$, and the coalescence end time, \TEND{}.
The \CHIEFF{} is a mass-weighted combination of the component spins parallel to the orbital angular momentum $\hat{L}$, defined as: 
\begin{equation}
\CHIEFF = \frac{(m_1 s_{1,z} + m_2 s_{2,z})\cdot \hat{L}}{m_1 + m_2},
\end{equation}
where we take $\hat{L}$ to be in the $z$-direction.
Both the injected and recovered masses quoted in this paper are in the detector frame.
The error on a recovered parameter, $\lambda$ is defined as: 
\begin{equation}
\label{eq:param-acc}
\mathrm{error} = \frac{\mathrm{recovered}~\lambda - \mathrm{injected}~\lambda}{\mathrm{injected}~\lambda},
\end{equation}
for all parameters except for the end time, where we simply take the error as the difference between the recovered and injected end times in milliseconds. 

\begin{figure}
\includegraphics[scale=0.5]{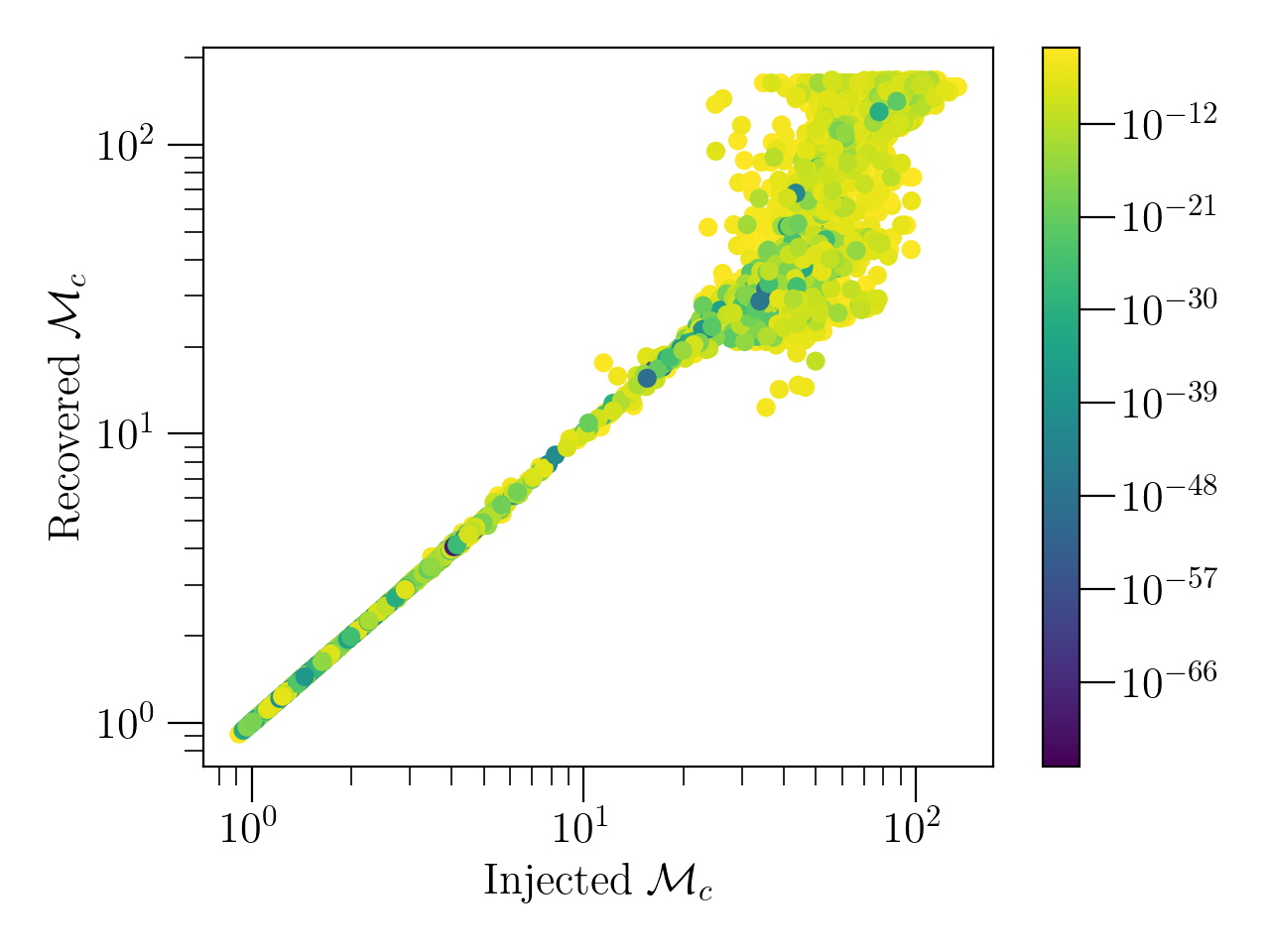}
\caption{\label{fig:mchirp-error-scatter}
Injected \MCHIRP{} for injections found with \ac{FAR} $<\TWOPERDAY{}$ is shown on the horizontal axis.
The vertical axis shows the recovered \MCHIRP{}.
The color bar is \ac{FAR}.
}
\end{figure}

\begin{figure}
\includegraphics[scale=0.5]{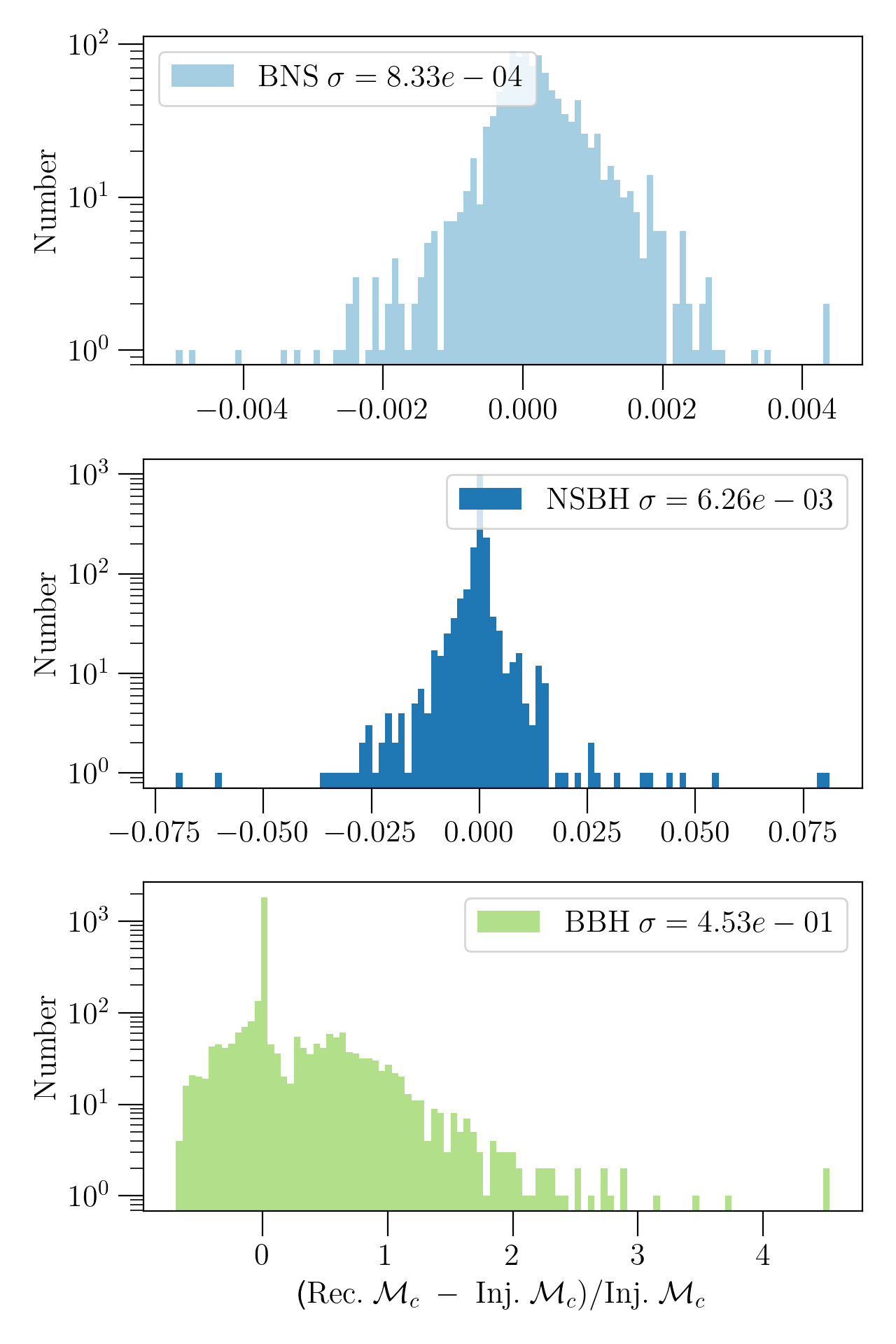}
\caption{\label{fig:mchirp-error-histogram}
Recovered \MCHIRP{} error for injections found with \ac{FAR} $<\TWOPERDAY{}$.
The top panel shows injections in the \ac{BNS} range of the parameter space, the middle panel shows \ac{NSBH} detections, and the bottom panel shows \ac{BBH} injections. 
The $\sigma$ value in each panel indicates the standard deviation on the recovered \MCHIRP{} error.
}
\end{figure}

It is well known that the chirp mass, \MCHIRP{}, is one of the best measured parameters in gravitational wave detections, however the recovered accuracy is highly dependent upon the mass of the system. 
Injections with small \MCHIRP $ < 10 - 20 \MSUN$ are recovered with very accurate \MCHIRP, but above this level, the accuracy starts to fall off, as shown in ~\figref{fig:mchirp-error-scatter}. 
For \ac{BNS} injections, we find a mean \MCHIRP{} error \BNSMCHIRPMEAN{} with a standard deviation of \BNSMCHIRPSTDEV{}.
Similarly, the \MCHIRP in the \ac{NSBH} region is recovered very well with mean \NSBHMCHIRPMEAN{} and standard deviation \NSBHMCHIRPSTDEV{}.
The \ac{BBH} region has a higher \MCHIRP{} error over all, and additionally a much larger spread in the error with mean \BBHMCHIRPMEAN{} and standard deviation \BBHMCHIRPSTDEV{}. 
Histograms of the recovered \MCHIRP{} error for each source class are shown in ~\figref{fig:mchirp-error-histogram}.

\begin{figure}
\includegraphics[scale=0.5]{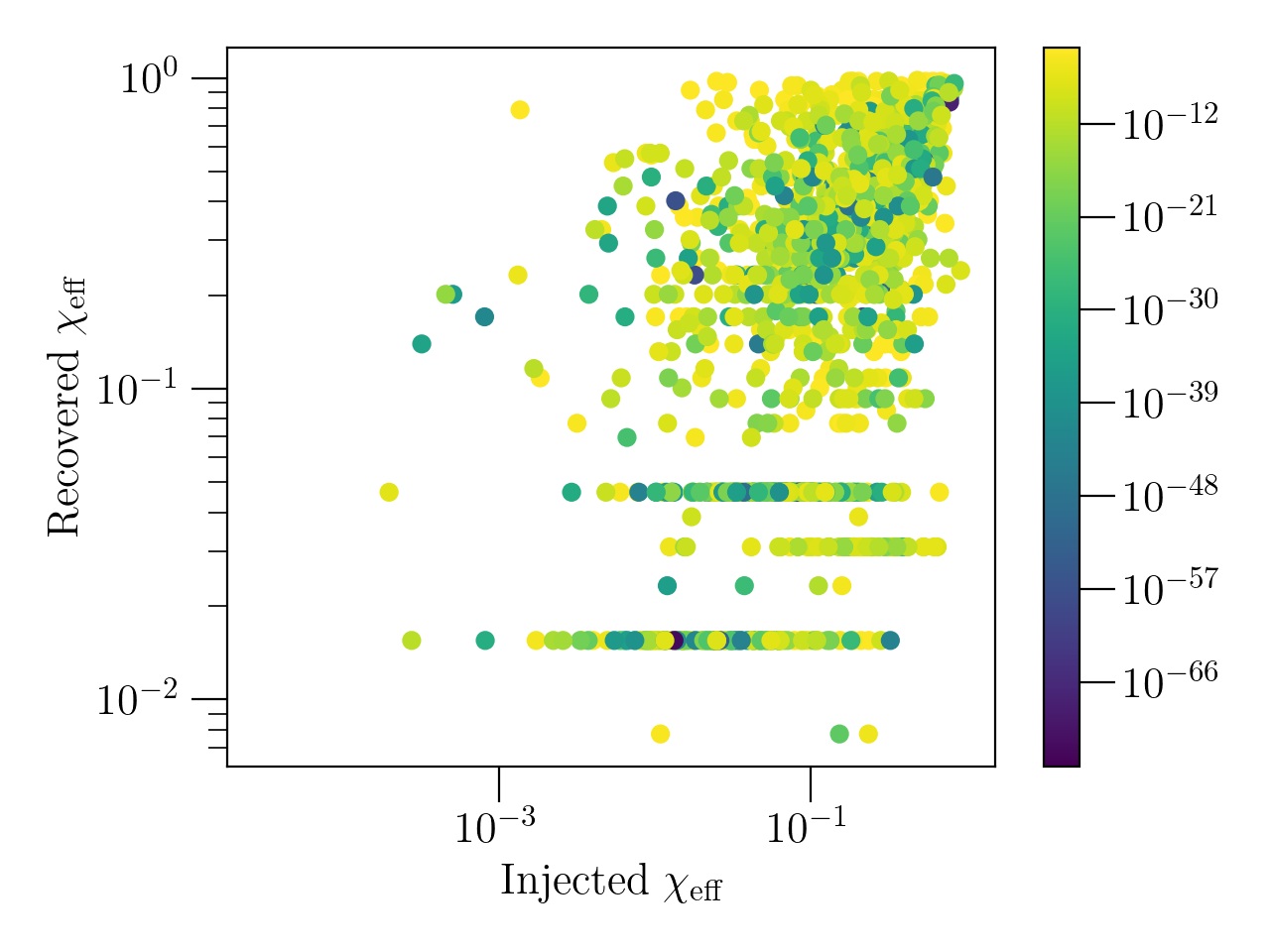}
\caption{\label{fig:chieff-error-scatter}
Injected \CHIEFF{} for injections found with \ac{FAR} $<\TWOPERDAY{}$ is shown on the horizontal axis.
The vertical axis shows the recovered \CHIEFF{}.
The color bar is \ac{FAR}.
}
\end{figure}

\begin{figure}
\includegraphics[scale=0.5]{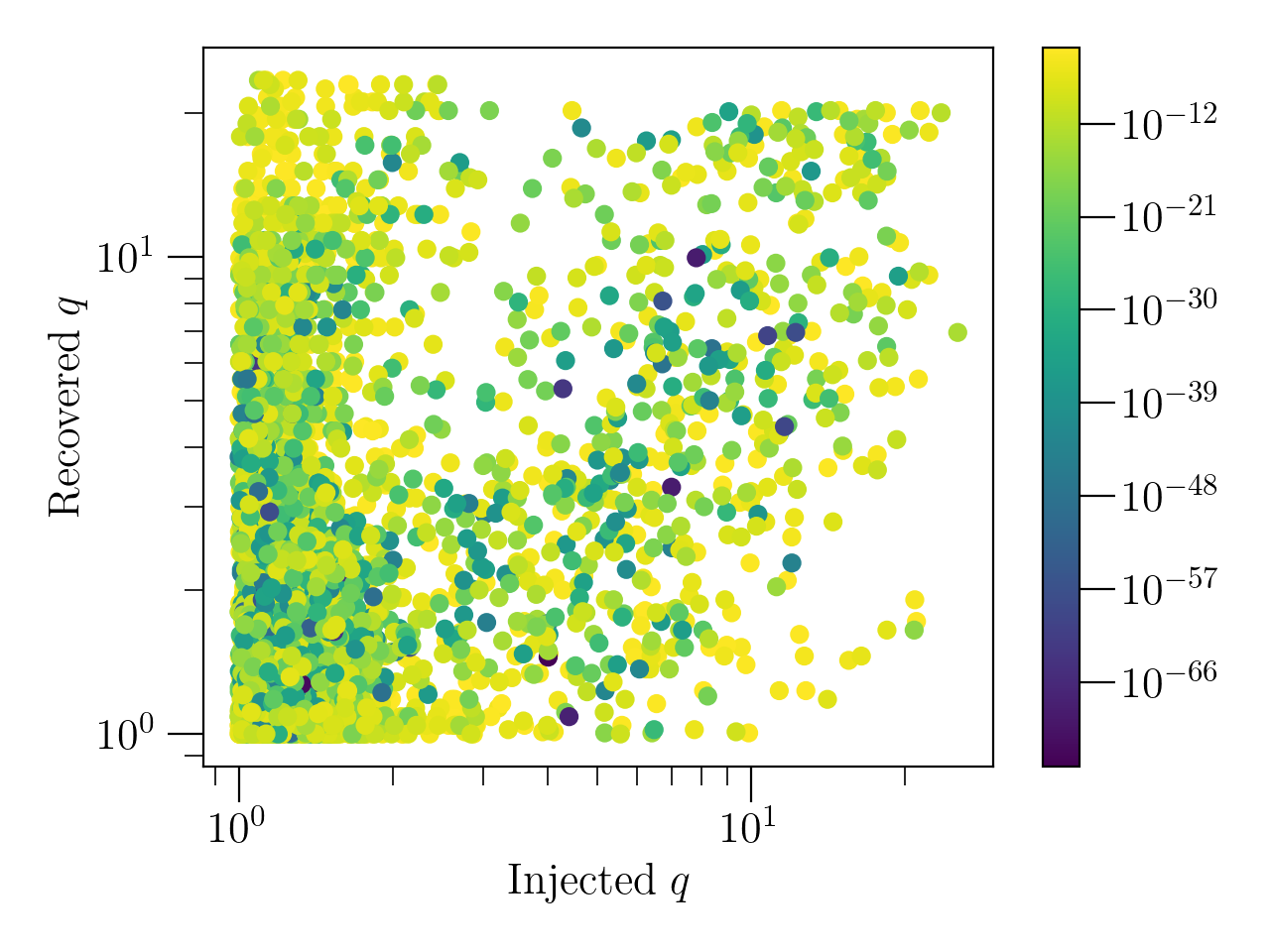}
\caption{\label{fig:q-error-scatter}
Injected mass ratio, $q = m_1 / m_2$, for injections found with \ac{FAR} $<\TWOPERDAY{}$ is shown on the horizontal axis.
The vertical axis shows the recovered mass ratio.
The color bar is \ac{FAR}.
}
\end{figure}

\begin{figure}
\includegraphics[scale=0.5]{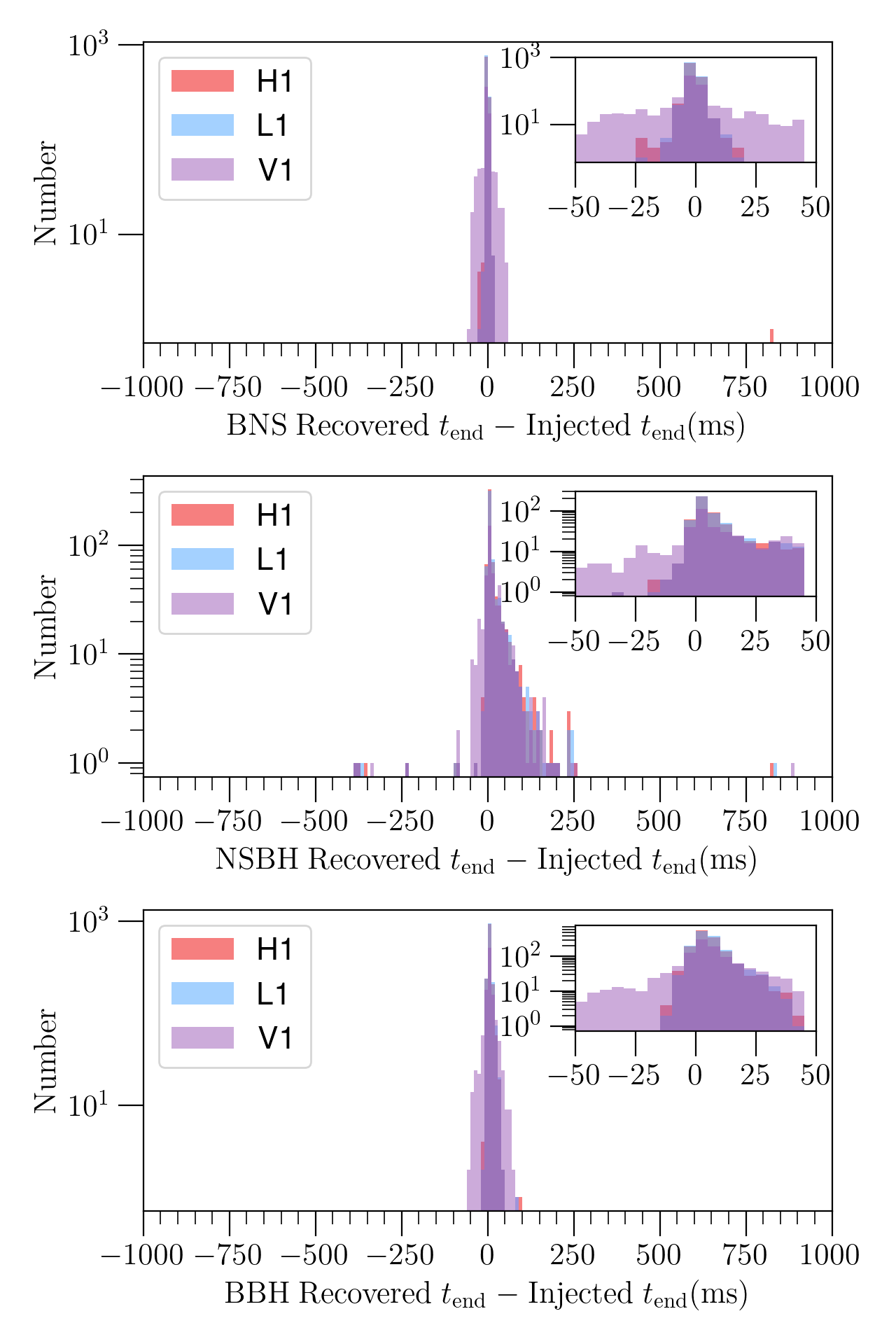}
\caption{\label{fig:end-time-hist}
Recovered end time accuracy in milliseconds of injections recovered with \ac{FAR} $<\TWOPERDAY{}$.
Results are shown for each interferometer: LIGO Hanford (red), LIGO Livingston (blue), and Virgo (purple). 
\ac{BNS} injections are shown in the upper panel, \ac{NSBH} in the center, and \ac{BBH} in the lower panel. 
}
\end{figure}

\figref{fig:chieff-error-scatter} and ~\figref{fig:q-error-scatter} are scatter plots of the injected and recovered \CHIEFF{} and $q$ respectively. 
These plots show that there is very little correlation between the injected and recovered values of these parameters. 
The mean and standard deviation on the recovered error for these parameters are given in Table~\ref{tab:param-acc}. 

A histogram of the difference between the injected and recovered injection end times is given in ~\figref{fig:end-time-hist}. 
Table~\ref{tab:param-acc} shows the mean \TEND{} difference across all source classes and detectors is \MEAN{ENDTIME} milliseconds with a standard deviation of \STDEV{ENDTIME} milliseconds.
The $90^{th}$ percentile on $|\TEND{}|$ is \QNINETY{ENDTIME} milliseconds and we recover every injection with a recovered \TEND{} less than a second away from the injected value.
~\figref{fig:end-time-hist} shows that for \ac{BNS} and \ac{BBH} injections, most of the recovered end times fall within $\pm 50$ milliseconds of the true injected end time. 
For \ac{BNS} injections, the mean \TEND{} difference is \BNSENDTIMEMEAN{} milliseconds with a standard deviation \BNSENDTIMESTDEV{}.
And for \ac{BBH} injections, the mean \TEND{} difference is \BBHENDTIMEMEAN{} milliseconds with a standard deviation \BBHENDTIMESTDEV{}.
However, for \ac{NSBH} templates, there is a wider distribution, skewing towards higher positive values of the end time difference, with mean \NSBHENDTIMEMEAN{} milliseconds and standard deviation \NSBHENDTIMESTDEV{}.
This is expected to be due to waveform systematics. 

\begin{event_table}
\begin{table}
{\small
\noindent\begin{tabularx}{\columnwidth}{l@{\extracolsep{\fill}} r r r r r }
& $\boldsymbol{\bar{X}}$ & $\boldsymbol{\sigma}$ & $\boldsymbol{P_{50}} $& $\boldsymbol{P_{75}}$ & $\boldsymbol{P_{90}}$ \\
\hline
\makebox[0pt][l]{\fboxsep0pt\colorbox{lightgray}{\mystrut\hspace*{1.0\linewidth}}} \MCHIRP{} & \MEAN{MCHIRP} & \STDEV{MCHIRP} & \QFIFTY{MCHIRP} & \QSEVENTYFIVE{MCHIRP} & \QNINETY{MCHIRP}  \\
\CHIEFF & \MEAN{CHIEFF} & \STDEV{CHIEFF} & \QFIFTY{CHIEFF} & \QSEVENTYFIVE{CHIEFF} & \QNINETY{CHIEFF}  \\
\makebox[0pt][l]{\fboxsep0pt\colorbox{lightgray}{\mystrut\hspace*{1.0\linewidth}}} $q$ & \MEAN{MASSRATIO} & \STDEV{MASSRATIO} & \QFIFTY{MASSRATIO} & \QSEVENTYFIVE{MASSRATIO} & \QNINETY{MASSRATIO}  \\
$t_{\mathrm{end}}$ & \MEAN{ENDTIME} & \STDEV{ENDTIME} & \QFIFTY{ENDTIME} & \QSEVENTYFIVE{ENDTIME} & \QNINETY{ENDTIME}  \\
\hline
\end{tabularx}
}
\caption{
Mean, $\bar{X}$, standard deviation, $\sigma$, and the fiftieth, seventy-fifth, and ninetieth percentiles on the recovered parameter error.
The error is defined as in Eqn.~\ref{eq:param-acc} for all parameters except for the end time, where we simply take the difference in milliseconds between the recovered and injected values as the error.
Results are computed only including injections which were recovered below a \ac{FAR} threshold of two per day.
The percentiles are computed for the absolute value of each distribution. 
}
\label{tab:param-acc}
\end{table}
\end{event_table}

\subsubsection{Search sensitivity}

\begin{figure}[h]
\includegraphics[scale=0.5]{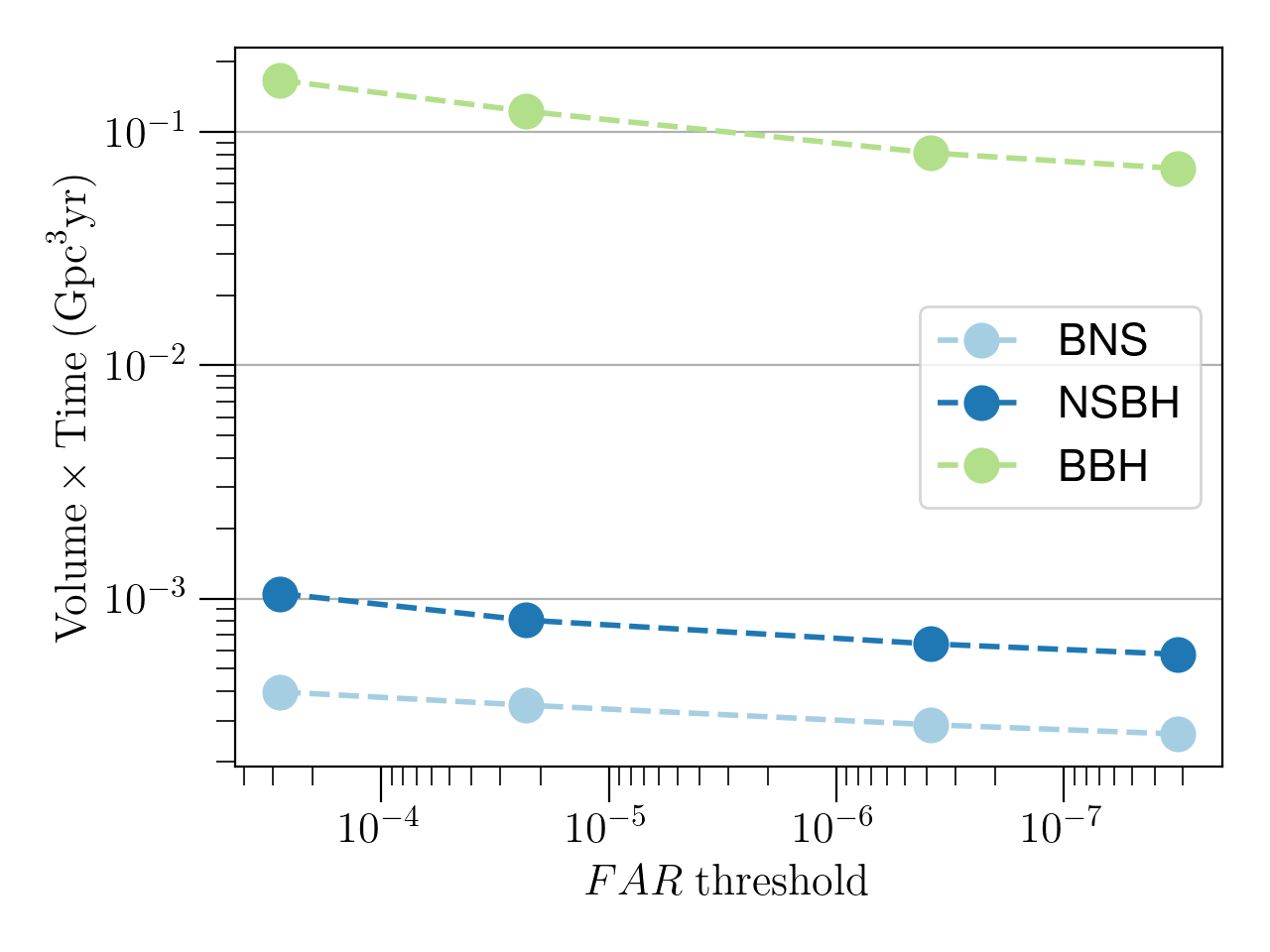}
\caption{\label{fig:vt}
\VT{} in each source class at the end of the \ac{MDC} at four different \ac{FAR} thresholds: $2$ per year, $1$ per month, \LOWFARTHRESH{}, and \HIGHFARTHRESH{}. 
\ac{BNS} \VT{} is shown in light blue, \ac{NSBH} \VT{} in dark blue, and \ac{BBH} \VT{} in green.
}
\end{figure}

\begin{figure}[h]
\includegraphics[scale=0.5]{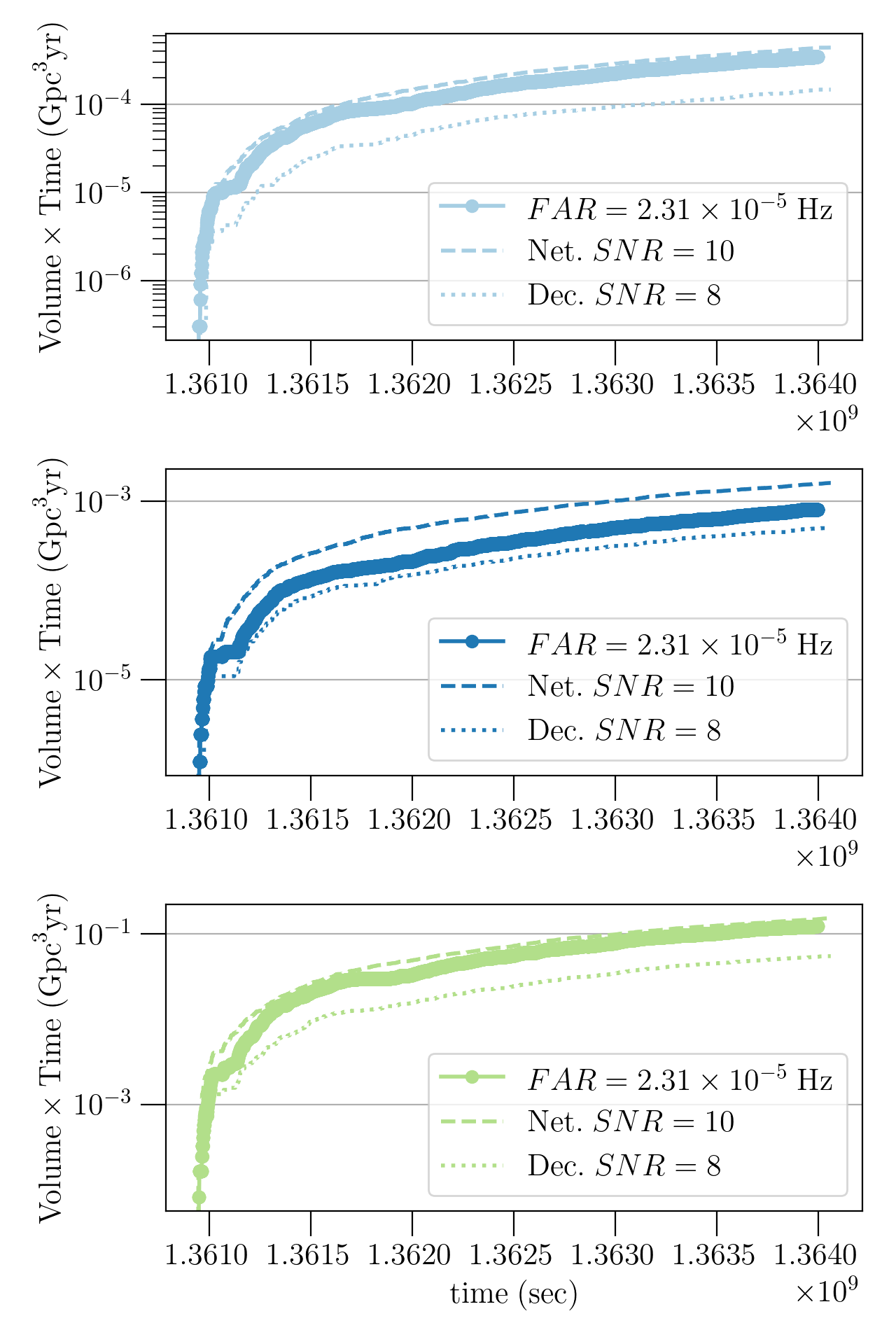}
\caption{\label{fig:vt-timeseries}
The plot shows the cumulative \VT{} time-series over the duration of the \ac{MDC} for each source class: \ac{BNS} in light blue (top panel), \ac{NSBH} in dark blue (middle panel), and \ac{BBH} in green (bottom panel).
The \VT{} is calculated using three different thresholds for counting ``found" injections: \ac{FAR} $ < \TWOPERDAY{}$ (dot markers), network \ac{SNR} $< \NETWORKSNRTHRESH{}$, and decisive \ac{SNR} $> \DECISIVESNRTHRESH{}$ (dotted line).
}
\end{figure}

The comoving volume is defined as~\cite{Hogg:1999ad,Chen:2017wpg}:
\begin{equation}
V_c = 4\pi D_H \int_0^z \mathrm{d}z \frac{(1+z)D_A^2}{E(z)}
\end{equation}
with $z$ as the redshift. 
$D_H$ is the Hubble distance, $D_A$ is angular distance, and $E(z)$ is the Hubble parameter. 
Multiplying this quantity by an observation time gives the surveyed spacetime-volume. 
We compute the total injected volume-time, $\VT{}_{\mathrm{inj}}$, using the max redshift to which injections were distributed and the time range over which injections were placed.
For the \ac{MDC} this time range is \MDCDURATION{} seconds.
With this quantity we can then estimate the online \VT{} in the \ac{MDC} as:
\begin{equation}
\langle VT \rangle = N_{f} \times \langle VT\rangle_{\mathrm{inj}}.
\end{equation}
Here, $N_f$ is the fraction of ``found" injections out of the total number of injections in the data. 
We independently compute the \VT{} for each source population.
The max redshifts of the injection distributions are $z =~\BNSMAXZ{}$, \NSBHMAXZ{}, and \BBHMAXZ{} for \ac{BNS}, \ac{NSBH} and \ac{BBH} respectively. 
Table~\ref{tab:vt} gives the injected \VT{} for each source class. 

\begin{event_table}
\begin{table}
{\small
\noindent\begin{tabularx}{\columnwidth}{l@{\extracolsep{\fill}} r r r }
$\boldsymbol{\VT{}}~\GPCYRS$ & \textbf{\ac{BNS}} & \textbf{\ac{NSBH}} & \textbf{\ac{BBH}} \\
\hline
\makebox[0pt][l]{\fboxsep0pt\colorbox{lightgray}{\mystrut\hspace*{1.0\linewidth}}} $\VT{}_{\mathrm{inj}}$ & \INJECTEDVT{BNS}& \INJECTEDVT{NSBH} & \INJECTEDVT{BBH}  \\
$\VT{}$ & \VTTWOPERDAY{BNS} & \VTTWOPERDAY{NSBH} & \VTTWOPERDAY{BBH}  \\
\hline
\end{tabularx}
}
\caption{
Values of the \VT{} in cubic gigaparsec-years measured at the end of the \ac{MDC} using a \ac{FAR} threshold of \TWOPERDAY{} compared to the injected \VT{} in each source class. 
}
\label{tab:vt}
\end{table}
\end{event_table}

~\figref{fig:vt} shows the final \VT{} in the \ac{MDC} for each source class using four different \ac{FAR} thresholds to determine whether injections count as found. 
In ~\figref{fig:vt-timeseries}, we show the cumulative \VT{} over the duration of the \ac{MDC} using different thresholds to count found injections.
For each source class, we find the highest \VT{} by using a threshold of network \ac{SNR} $=\NETWORKSNRTHRESH{}$ and the lowest \VT{} with a threshold of decisive \ac{SNR} $=\DECISIVESNRTHRESH{}$. 

\subsubsection{Sky localization and source classification}
\label{sec:skymap-pastro}

\begin{figure}[h]
\includegraphics[scale=0.5]{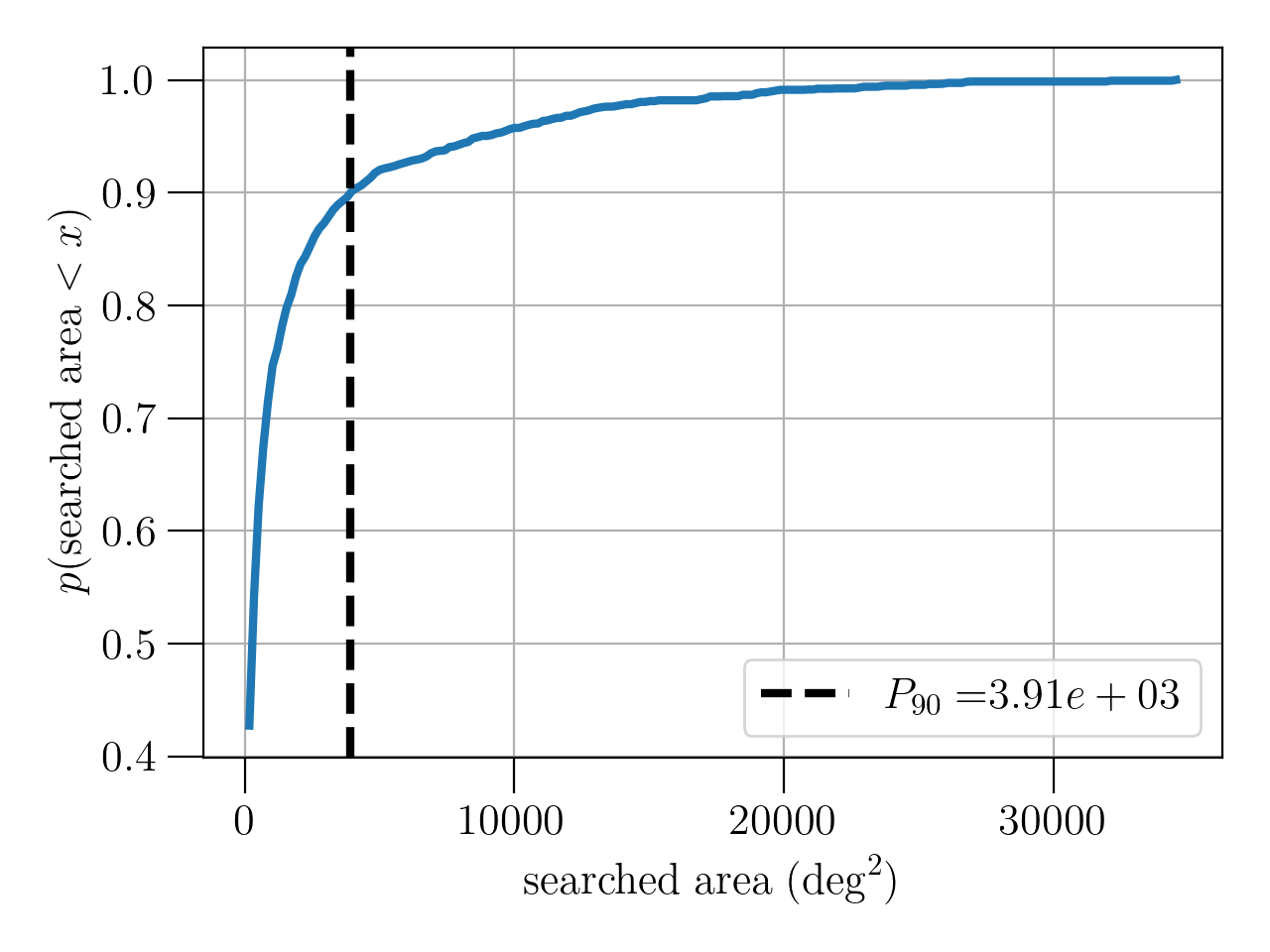}
\caption{\label{fig:searched-area-twoperday}
Cumulative distribution function of skymap searched area for injections recovered with \ac{FAR} less than two per day.
}
\end{figure}

\begin{figure}[h]
\includegraphics[scale=0.5]{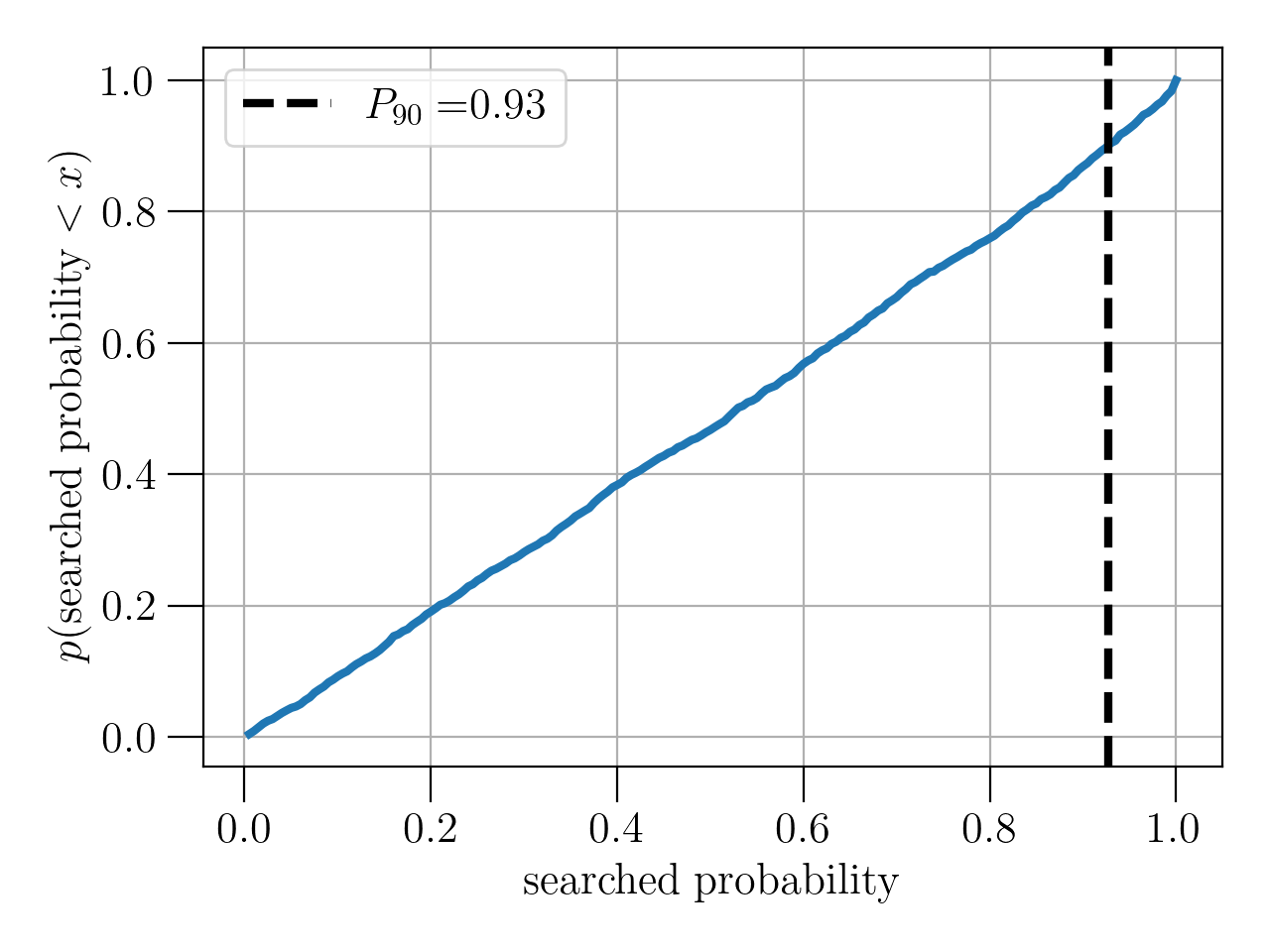}
\caption{\label{fig:searched-prob-twoperday}
Cumulative distribution function of skymap searched probability for injections recovered with \ac{FAR} less than two per day.
}
\end{figure}

For efficient electromagnetic follow-up, it is vital that accurate sky localizations and source classifications are provided to the public in low latency. 
The sky localization information informs where electromagnetic observers should search on the sky to find coincident events. 
The accuracy and precision of sky localization information can have a direct impact on the time it takes to identify a counterpart, especially for narrow field of view telescopes. 
Sky localizations are produced in low latency for all events on \ac{GRACEDB} using Bayestar~\cite{Singer:2015ema,Singer:2016eax}. 
The sky localization calculation depends on the \ac{SNR} time-series around the coalescence time of the event.
These are uploaded to \ac{GRACEDB} by the search pipelines as part of the event metadata. 
Since the potential for bright electromagnetic counterparts is highly dependent on the nature of the binary source, it is also important to provide accurate source classification so that observers may make informed decisions about when to follow up gravitational wave events. 
The probability that a gravitational wave candidate is astrophysical in origin is the \PASTRO{}, computed by \GSTLAL{} using the multi-component FGMC formalism~\cite{Farr:2013yna,Kapadia:2019uut,Ray:2023nhx}. 
We use a population model with a Salpeter distribution for the source component masses $m_1$, $m_2$ given by~\cite{Salpeter:1955it}:
\begin{equation}
\label{eq:pop-model}
p(m_1, m_2) \propto \frac{m_{1}^{-2.35}}{m_1 - m_{\mathrm{min}}}, 
\end{equation}
with $m_{\mathrm{min}} = 0.8\MSUN{}$ and a uniform distribution in component spins, $s_{1,z}$, $s_{2,z}$. 
The \PASTRO{} is further divided into the probability that a gravitational wave candidate originates from a \ac{BNS}, \ac{NSBH}, or \ac{BBH} as: 
\begin{equation}
\begin{split}
p(\mathrm{astro}) & = 1 - p(\mathrm{Terrestrial}) \\
& = p(\ac{BNS}) + p(\ac{NSBH}) + p(\ac{BBH}).
\end{split}\label{popmodel}
\end{equation}
For this purpose, we use a cutoff of \NSMASSHIGH{} as the maximum neutron star mass to define the \ac{BNS}, \ac{NSBH}, and \ac{BBH} regions.
In this section, we will briefly summarize the accuracy of the Bayestar skymaps as well as the FGMC \PASTRO{} for injections recovered by \GSTLAL{} during the \ac{MDC}. 
Detailed information about recent developments to the FGMC \PASTRO{} calculation and a comparison between offline and online \PASTRO{} results for \GSTLAL{} events is given in~\cite{Ray:2023nhx}.

To quantify the sky localization performance we will consider the skymap searched area and searched probability. 
The searched area is given in square degrees and represents the area within the credible region containing the injection's true source location. 
This gives an indication of the accuracy of the skymaps. 
The searched area can be interpreted as the sky area an electromagnetic observer would have to tile before reaching the true injection sky location assuming they start at the highest probability sky position given by the skymap. 
Therefore, a smaller searched area is desirable as it means an observer could find the injection with minimal telescope pointings. 
The searched probability is similarly the probability within the credible region containing the injection's true sky location. 
We aim to find $90\%$ of injections within the $90\%$ credible region of the true source location. 
If the searched probability P-P plot lies off the diagonal, we can make the interpretation that there is an inconsistency in the sky localizations, i.e. they may be accurate but lack precision, or vice versa. 

\begin{event_table}
\begin{table}
{\small
\noindent\begin{tabularx}{\columnwidth}{l@{\extracolsep{\fill}} r r r }
 & $\boldsymbol{P_{50}}$ & $\boldsymbol{P_{75}}$ & $\boldsymbol{P_{90}}$ \\
\hline
\makebox[0pt][l]{\fboxsep0pt\colorbox{lightgray}{\mystrut\hspace*{1.0\linewidth}}} ALL & \SEARCHEDAREAQFIFTY{ALL} & \SEARCHEDAREAQSEVENTYFIVE{ALL} & \SEARCHEDAREAQNINETY{ALL}  \\
$1$ IFO & \SEARCHEDAREAQFIFTY{SINGLE} & \SEARCHEDAREAQSEVENTYFIVE{SINGLE} & \SEARCHEDAREAQNINETY{SINGLE}  \\ 
\makebox[0pt][l]{\fboxsep0pt\colorbox{lightgray}{\mystrut\hspace*{1.0\linewidth}}} $2$ IFO & \SEARCHEDAREAQFIFTY{DOUBLE} & \SEARCHEDAREAQSEVENTYFIVE{DOUBLE} & \SEARCHEDAREAQNINETY{DOUBLE}  \\
$3$ IFO & \SEARCHEDAREAQFIFTY{TRIPLE} & \SEARCHEDAREAQSEVENTYFIVE{TRIPLE} & \SEARCHEDAREAQNINETY{TRIPLE}  \\
\hline
\end{tabularx}
}
\caption{
Fiftieth, seventy-fifth, and ninetieth percentiles on the searched area of injections of each coincidence type recovered with \ac{FAR} less than two per day. 
Values are given in $\mathrm{deg}^2$.
}
\label{tab:searched-area}
\end{table}
\end{event_table}

~\figref{fig:searched-area-twoperday} shows the cumulative distribution function of the searched area for all injections recovered by \GSTLAL{} with \ac{FAR} less than two per day. 
We find that the $90^{th}$ percentile of searched area is $\SEARCHEDAREAQNINETY{ALL}~\mathrm{deg}^2$. 
This is about $19\%$ of one hemisphere of the sky. 
Since we know that injections recovered in coincidence by two or three interferometers will have more accurate sky localizations, we give statistics on the searched areas by coincidence type in Table~\ref{tab:searched-area}. 
~\figref{fig:searched-prob-twoperday} shows the cumulative distribution function of the searched probabilities. 
We find that the fiftieth, seventy-fifth, and ninetieth percentiles on the searched probability are \SEARCHEDPROBQFIFTY{ALL}, \SEARCHEDPROBQSEVENTYFIVE{ALL}, and \SEARCHEDPROBQNINETY{ALL} respectively. 
The CDF lies very near to the diagonal, indicating that Bayestar is producing well-calibrated sky localizations.

\begin{figure}[h]
\includegraphics[width=\columnwidth]{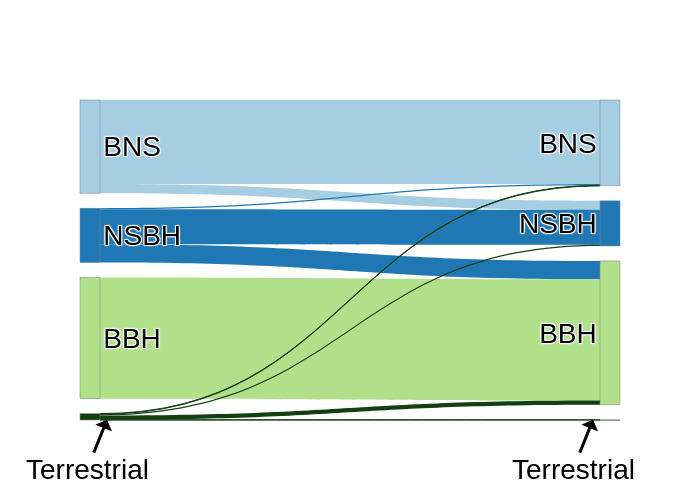}
\caption{\label{fig:pastro}
Sankey diagram showing \PASTRO{} classification of events uploaded to \ac{GRACEDB} during the \ac{MDC} with \ac{FAR} less than two per day. 
Events with an end time within $\pm 1$ second of an injection are classified as either \ac{BNS}, \ac{NSBH}, or \ac{BBH} using a neutron star mass boundary of \NSMASSHIGH{}. 
Events that do not correspond in time with an injection are all classified as terrestrial.
}
\end{figure} 

We will briefly discuss the performance of the \PASTRO{} in the \ac{MDC} using the Sankey diagram in ~\figref{fig:pastro}. 
The diagram is read from left to right. 
The width of each source band on the left corresponds to $N_{\mathrm{source}}$, the number of recovered events in each source class: \ac{BNS}, \ac{NSBH}, \ac{BBH}, and terrestrial. 
The terrestrial events are any candidates uploaded to \ac{GRACEDB} which do not coincide in time with an injection. 
The width of each band on the right is the sum of recovered $p(\mathrm{source})$. 
This diagram gives an indication of the relative misclassification between sources. 

For \ac{BNS} injections, we find that $p(\ac{BNS})$ accounts for \BNSTOBNS{} of the recovered probabilities, while $p(\ac{NSBH})$ accounts for \BNSTONSBH{}. 
The recovered \ac{BBH} and terrestrial probabilities of \ac{BNS} injections are negligible. 
This indicates that \ac{BNS} signals were most commonly mistaken as \ac{NSBH}.
For \ac{NSBH} injections, we find $p(\ac{NSBH})$ is \NSBHTONSBH{} of the recovered \PASTRO{}, $p(\ac{BBH})$ makes up \NSBHTOBBH{}, and $p(\ac{BNS})$ makes up \NSBHTOBNS{}. 
There is a significant amount of misclassification between \ac{BNS} and \ac{NSBH} signals.
However, the misclassification is asymmetrical with very few \ac{NSBH} injections being assigned high $p(\ac{BNS})$ whereas many more \ac{BNS} injections are assigned a high $p(\ac{NSBH})$. 
This misclassification is an ongoing area of study and corrections to the \PASTRO{} calculation which mitigate this effect are discussed in~\cite{Ray:2023nhx}. 
The \PASTRO{} calculation performs very well for \ac{BBH} classification, with \BBHTOBBH{} recovered $p(\ac{BBH})$.

\subsubsection{Latency performance}
\label{sec:latency}

\begin{figure}[h]
\includegraphics[scale=0.5]{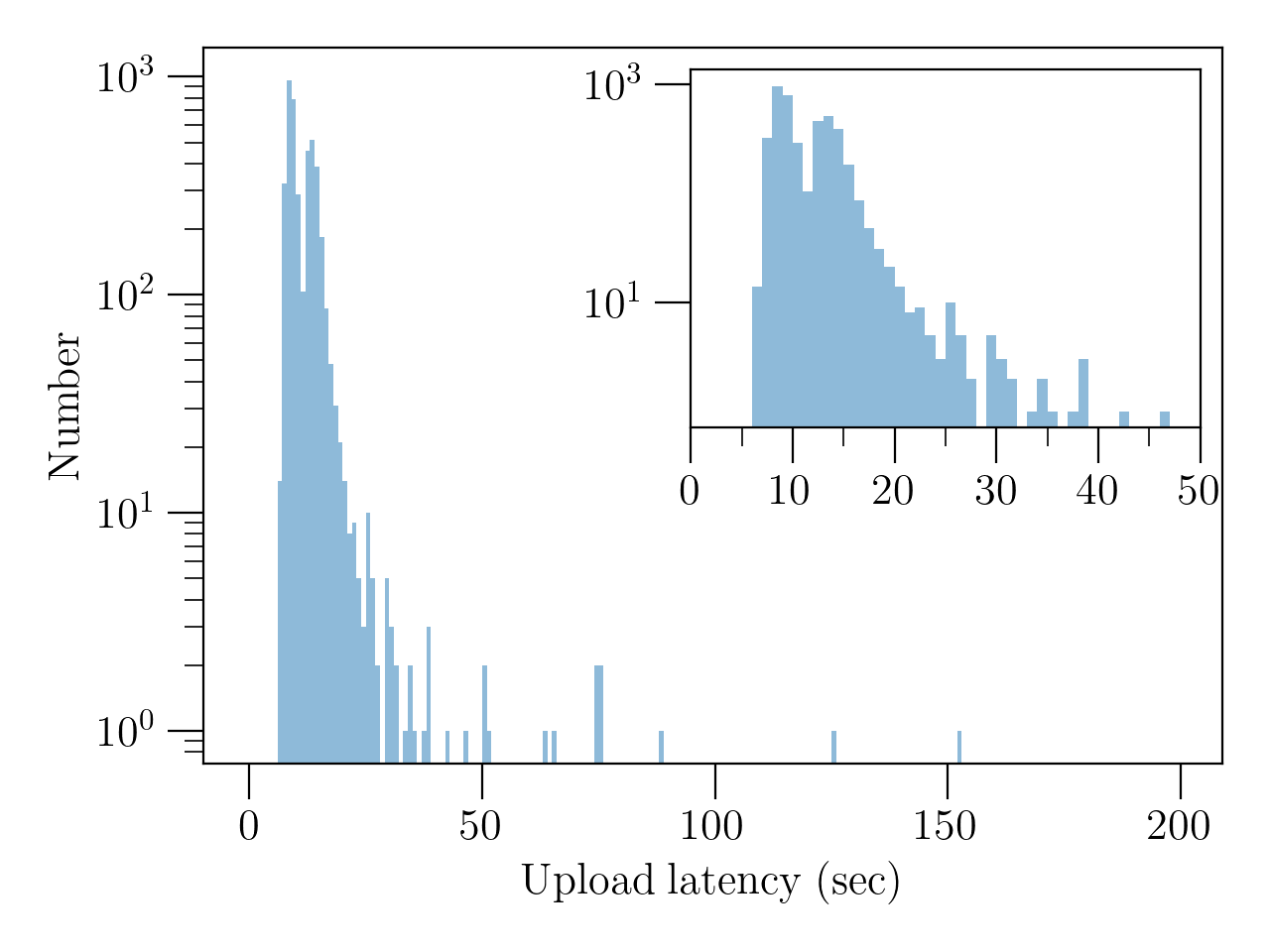}
\caption{\label{fig:latency}
Upload latency, defined as the difference between the GPS time of upload to \ac{GRACEDB} and the event coalescence time, in seconds. 
}
\end{figure}

~\figref{fig:latency} shows a histogram of the upload latency for all recovered injections on \ac{GRACEDB}. 
The upload latency is defined as the difference between the time the event appears on \ac{GRACEDB} and the event coalescence time.
The distribution is bimodal as a result of the built-in \UPLOADCADENCE{} second geometric wait time between uploads, discussed in Sec.~\ref{sec:gstlal}. 
The first peak in the upload latency distribution is at $8-9$ seconds and the second peak is at $13-14$ seconds. 
This shows that the \GSTLAL{} pipeline is able to keep up with filtering data in real time and regularly produce gravitational wave candidates within $\mathcal{O}(10)$ seconds of the coalescence time. 

\begin{figure}[h]
\includegraphics[scale=0.5]{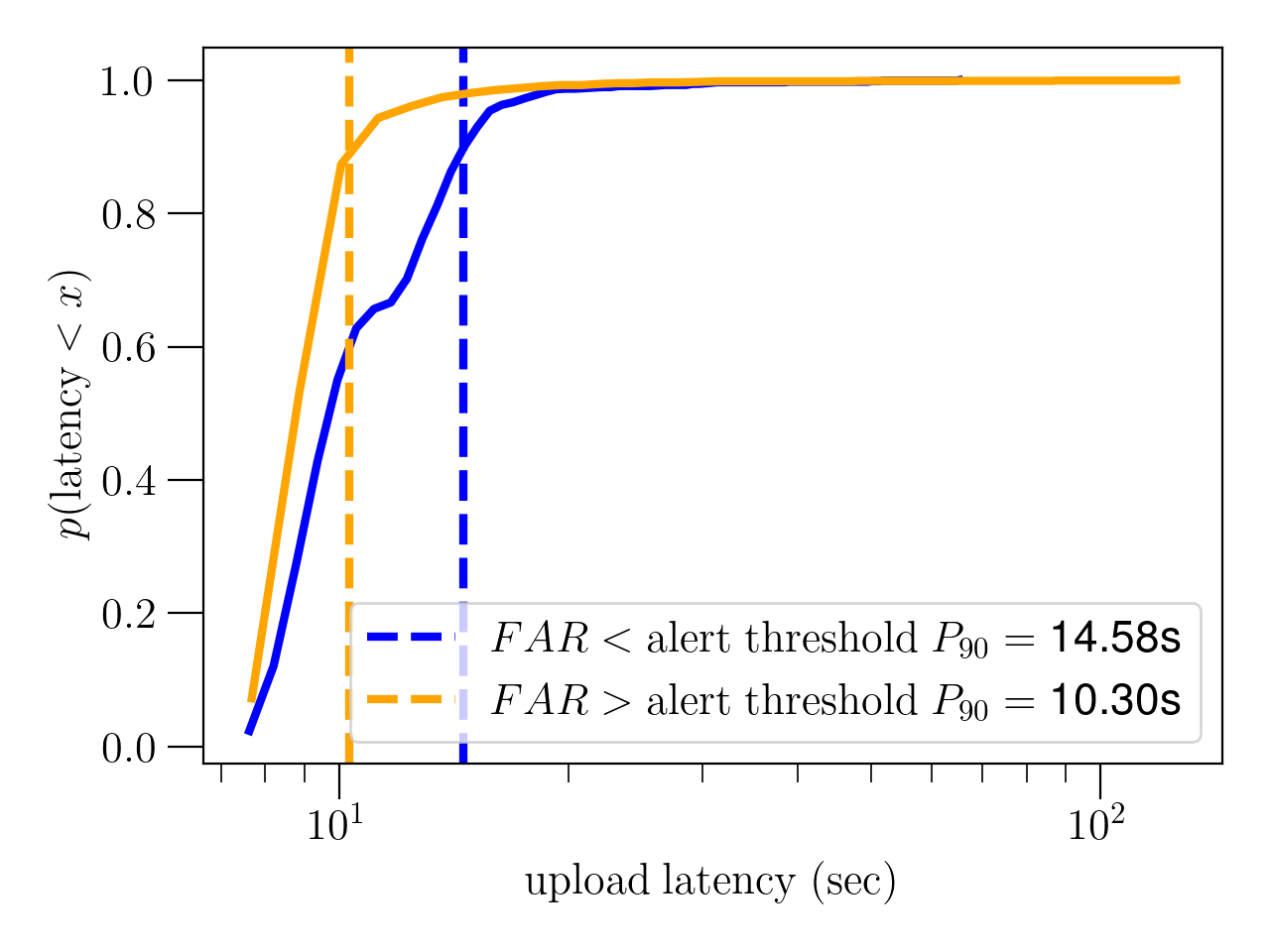}
\caption{\label{fig:latency-cdf}
Cumulative distribution of event upload latencies. 
The orange (blue) curve shows the distribution of latencies for the first event uploaded with \ac{FAR} higher (lower) than the public alert threshold. 
The dashed lines show the location of the $90^{th}$ percentile for each distribution. 
}
\end{figure}

The \UPLOADCADENCE{} second wait time in the event aggregation is implemented in order to reduce the total number of uploads by waiting a long enough time to collect many triggers across background bins before making an upload. 
Even though there is no built-in wait time for the first event upload, subsequent uploads are necessarily delayed by this method. 
If an event with \ac{FAR} below the alert threshold comes after the first upload it can be delayed by at least \UPLOADCADENCE{} seconds. 
~\figref{fig:latency-cdf} shows the cumulative distribution of upload latency for the first events above and below the alert threshold for each superevent in the \ac{MDC}. 
We find that the low \ac{FAR} event uploads are significantly delayed by the event aggregation process.
After identifying this issue in the \ac{MDC} we plan to make changes to the aggregation scheme in order to reduce the latency of alert quality uploads before the start of \ac{O4}.

\section{Conclusion}
\label{sec:conclusion}

\GSTLAL{} is a matched-filtering based gravitational wave search pipeline, which is operated in a low-latency configuration in order to identify signals within seconds of their arrival. 
We have introduced the \ac{GWLTS} software as a useful auxiliary tool for characterizing the performance of such an analysis in real time. 
We have presented the performance of the \GSTLAL{} pipeline on a mock data challenge consisting of $40$ days \ac{H1}\ac{L1}\ac{V1} data from \ac{O3} along with an injection campaign of simulated \ac{BNS}, \ac{NSBH}, and \ac{BBH} signals. 

Within the \ac{MDC} data we recover $9$ previously published gravitational wave candidates at the one per hour \ac{FAR} threshold. 
As only five of these were identified by the \GSTLAL{} pipeline in low-latency in \ac{O3}, we have demonstrated an improvement in the pipeline's signal recovery.
We attribute this improvement to several incremental updates to the likelihood ratio computation~\cite{Tsukada:2023edh} and the new method of removing signals from the background as introduced in~\cite{Joshi:2023ltf}.
During the \ac{MDC}, we find only one candidate which, if uploaded during \ac{O4}, would likely be retracted. 
The candidate is identified in only a single detector and we expect that increasing the penalty applied to single detector candidates in the likelihood ratio would reduce our recovery of such spurious signals in the future. 
We have detailed the results of the injection campaign including efficiency of signal recovery across the parameter space, accuracy of estimated parameters, search sensitivity, sky localization and source classification accuracy, and typical latencies.  

The configuration and performance of the \GSTLAL{} pipeline as described in Sec.~\ref{sec:mdc} is a close approximation to what will be used in the fourth observing run of the \ac{LVK} Collaboration. 
However, since the conclusion of the \ac{MDC} used in this paper, work has been ongoing and several areas for possible improvements have been identified. 
These changes in configuration and the corresponding improvements in performance are given in Appendix~\ref{sec:appendix}.

\begin{acknowledgments}
    The authors are grateful for computational resources provided by the LIGO
    Laboratory and supported by National Science Foundation Grants PHY-0757058
    and PHY-0823459.  This material is based upon work supported by NSF's LIGO
    Laboratory which is a major facility fully funded by the National Science
    Foundation.  LIGO was constructed by the California Institute of Technology
    and Massachusetts Institute of Technology with funding from the National
    Science Foundation (NSF) and operates under cooperative agreement
    PHY-1764464. The authors gratefully acknowledge the Italian Istituto Nazionale 
    di Fisica Nucleare (INFN), the French Centre National de la Recherche Scientifique 
    (CNRS) and the Netherlands Organization for Scientific Research, for the construction 
    and operation of the Virgo detector and the creation and support  of the EGO consortium.  
    The authors are grateful for computational resources provided
    by the Pennsylvania State University's Institute for Computational and Data
    Sciences (ICDS) and the University of Wisconsin Milwaukee Nemo and support
    by NSF PHY-\(2011865\), NSF OAC-\(2103662\), NSF PHY-\(1626190\), NSF
    PHY-\(1700765\), and NSF PHY-\(2207728\).  This paper
    carries LIGO Document Number LIGO-P2300124. \\

\emph{Software:}
Confluent Kafka (\url{https://developer.confluent.io/get-started/python/}), Grafana (\url{https://grafana.com/docs/}), \ac{GRACEDB}~\cite{gracedb}, \GSTLAL{}~\cite{gstlal}, \texttt{GWCelery}~\cite{gwcelery}, \ac{GWLTS}~\cite{gwlts}, \IGWNALERT{}~\cite{igwn_alert}, InfluxDB (\url{https://docs.influxdata.com/influxdb/v2.7/}), \texttt{ligo-scald}~\cite{ligo_scald}

\appendix 

\section{\GSTLAL{} performance with updated \ac{O4} configuration}
\label{sec:appendix}

In this section, we demonstrate the performance of the \GSTLAL{} pipeline using an updated configuration. 
The improved performance in this updated run are attributed to the following areas of development, which proceeded after the conclusion of the \ac{MDC} presented in Sec.~\ref{sec:mdc}.

The analytic \ac{SNR}-$\xi^2$ signal model used in the likelihood ratio ranking statistic has now been tuned to use a more optimal allowed mismatch region, which is wider for the high \ac{SNR} region of parameter space. 
This change is expected to improve the recovery efficiency for very loud signals. 

The presence of non-Gaussian noise transients, known as glitches, in the strain data has long been a problem for gravitational wave searches. 
\ac{IDQ} is a machine-learning based algorithm used to assign probabilities of the presence of a glitch in a segment of strain data~\cite{Essick:2020qpo}. 
In the \ac{O3} offline analysis, \ac{IDQ} was incorporated into the \GSTLAL{} ranking statistic as a means of reducing the significance of candidates found during particularly glitchy stretches of data. 
Although it is no longer used in the ranking statistic, it is now possible to use \ac{IDQ} state information as a gate on the strain data, so that segments of data with a high glitch probability will be excluded from the filtering. 
This should mitigate the negative effects of non-Gaussian data and is expected to result in fewer retraction level candidates and an improved \VT{}. 
Although the \ac{IDQ} gate was used in the \ac{MDC} analysis presented here, the feature is still under development and is not planned for use in the \ac{O4} production configuration. 

As mentioned in Sec.~\ref{sec:skymap-pastro}, the \PASTRO{} performance in the \ac{BNS} and \ac{NSBH} region of the parameter space was sub-optimal in the \ac{MDC}, with a significant amount of misclassification between the two source types, as well as \ac{NSBH} misclassification as \ac{BBH}. 
It is imperative that sources including a neutron star are not falsely classified as \acp{BBH} since this may discourage astronomers from following up these potentially electromagnetically bright signals. 
Since the conclusion of the \ac{MDC} analysis presented in this paper, work has been ongoing toward improving the \PASTRO{} source classification.
This effort is discussed in more detail in~\cite{Ray:2023nhx}. 

Finally, as discussed in Sec.~\ref{sec:latency}, we have made improvements to the event aggregation method to ensure that events with \ac{FAR} below the public alert threshold will be uploaded with as little latency as possible. 
By adjusting the geometric cadence factor in the aggregator and removing the wait time for these low \ac{FAR} events we expect a reduction in the latency of these uploads by up to several seconds. 

We re-analyze the final two weeks of the \ac{MDC} data using the updated pipeline configuration in order to demonstrate the effect on the pipeline performance. 
The ``initial run" presented in Sec.~\ref{sec:mdc} was analyzed from 03-15-2023 to 03-28-2023 and the ``updated run"   was analyzed from 04-24-2023 to 05-07-2023. 
These runs correspond to identical stretches of \ac{O3} replay data with the same injections present in each. 

\figref{fig:snr-timeseries-mdc12} shows the decisive \ac{SNR} time-series for all injections within the two week span. 
There are several high \ac{SNR} injections missed by the initial analysis (upper panel) but found in the updated analysis (lower panel). 
This indicates an improvement as a result of the widened mismatch region of the \ac{SNR}-$\xi^2$ signal model. 
Table~\ref{tab:mdc12-efficiency} compares the injection efficiencies between the two runs. 
For the \ac{BNS} and \ac{NSBH} regions the efficiency is comparable, while there is a $6\%$ improvement in the \ac{BBH} efficiency from \BBHINBANKEFFICIENCY{TWOPERDAY} in the initial run to \BBHINBANKEFFICIENCYMDCTWELVE{TWOPERDAY} in the updated run. 

\begin{figure}[h]
\includegraphics[scale=0.35]{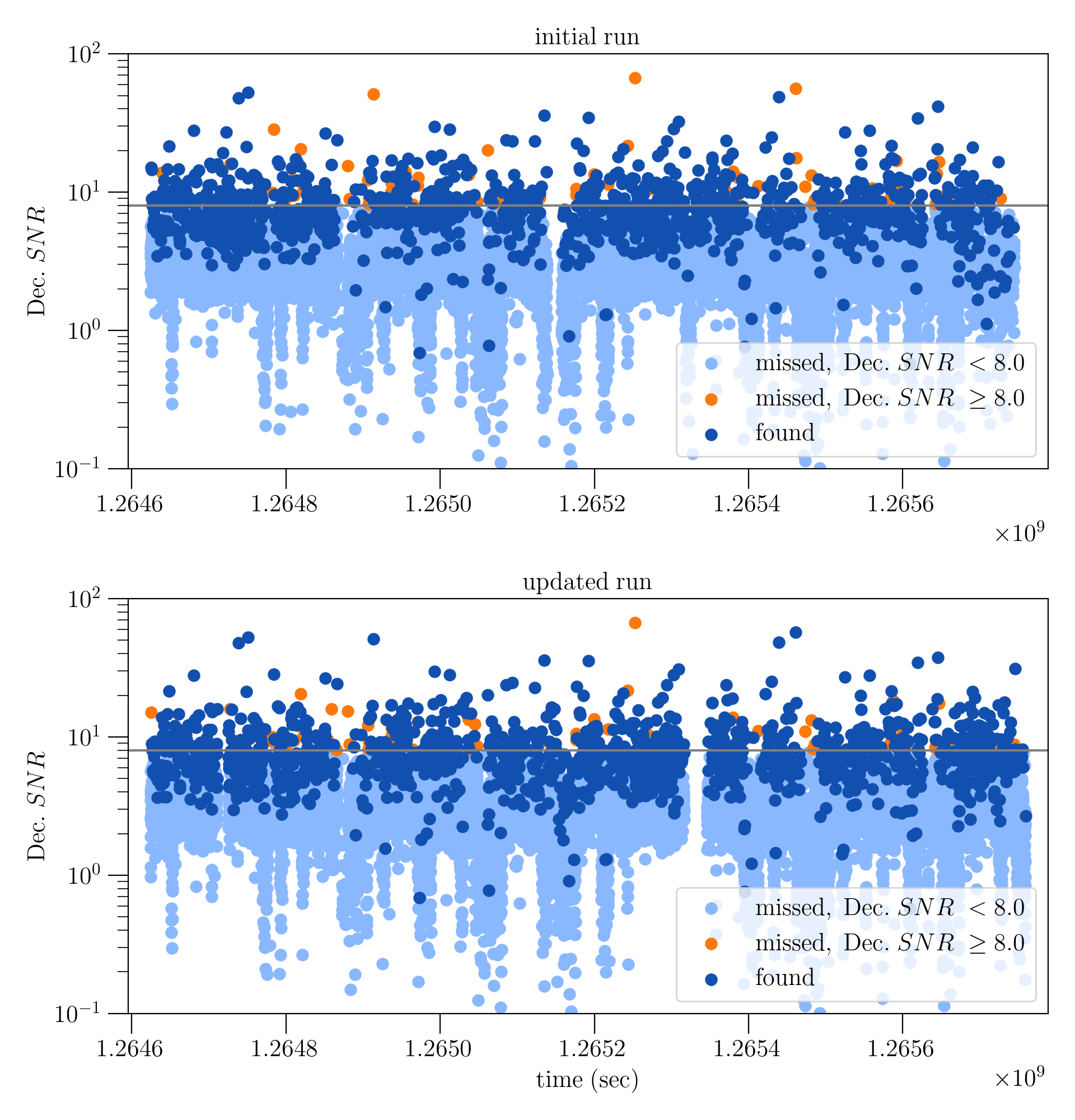}
\caption{\label{fig:snr-timeseries-mdc12}
Time-series of injected decisive \ac{SNR} for injections with component masses and spins within the \ac{O4} template bank in the initial run, 03-15-2023 to 03-28-2023 (upper panel) and updated run 04-24-2023 to 05-07-2023 (lower panel).
Dark blue markers indicate injections that were recovered below a \LOWFARTHRESH{} \ac{FAR} threshold. 
Orange and light blue markers indicate injections not recovered below this \ac{FAR} threshold, where orange points are injections with decisive \ac{SNR} $>\DECISIVESNRTHRESH{}$. 
Times on the horizontal axis are GPS times shifted to the original \ac{O3} epoch. 
}
\end{figure}

\begin{event_table}
\begin{table}
{\small
\noindent\begin{tabularx}{\columnwidth}{l@{\extracolsep{\fill}}c c c c}
  & \textbf{\ac{BNS}} & \textbf{\ac{NSBH}} & \textbf{\ac{BBH}} & \textbf{ALL} \\
\hline
\makebox[0pt][l]{\fboxsep0pt\colorbox{lightgray}{\mystrut\hspace*{1.0\linewidth}}} initial run & \BNSINBANKEFFICIENCY{TWOPERDAY} &\NSBHINBANKEFFICIENCY{TWOPERDAY} & \BBHINBANKEFFICIENCY{TWOPERDAY} & \ALLINBANKEFFICIENCY{TWOPERDAY} \\
updated run & \BNSINBANKEFFICIENCYMDCTWELVE{TWOPERDAY} &\NSBHINBANKEFFICIENCYMDCTWELVE{TWOPERDAY} & \BBHINBANKEFFICIENCYMDCTWELVE{TWOPERDAY} & \ALLINBANKEFFICIENCYMDCTWELVE{TWOPERDAY} \\
\hline
\end{tabularx}
}
\caption{
Injection efficiencies as defined in Eq.~\ref{eq:eff} computed using a \ac{FAR} threshold of \TWOPERDAY{} to count ``found" injections. 
The ``initial run" is the same as presented in Sec.~\ref{sec:mdc} and the ``updated run" uses the configuration improvements described in this section. 
Source categories are defined in Table~\ref{tab:in-bank} and ``ALL" combines injections from the three source categories. 
}
\label{tab:mdc12-efficiency}
\end{table}
\end{event_table}

\figref{fig:latency-hist-mdc12} and \figref{fig:latency-cdf-mdc12} show the improvements in event upload latency between the two runs. 
The histogram in~\figref{fig:latency-hist-mdc12} shows that upload latencies for the updated run are shifted slightly lower overall with respect to the initial run, while the lower edge remains the same. 
This is expected as the improvements focused on lowering the latency of subsequent uploads by reducing the geometric cadence factor from \UPLOADCADENCE{} seconds to \UPLOADCADENCEMDCTWELVE{} seconds. 
\figref{fig:latency-cdf-mdc12}  shows a similar improvement in upload latency for events below the public alert threshold. 
We find a $2.54$ second reduction in the $90^{th}$ percentile on upload latency for events with \ac{FAR} below the public alert threshold with respect to the initial run.

\begin{figure}[h]
\includegraphics[scale=0.5]{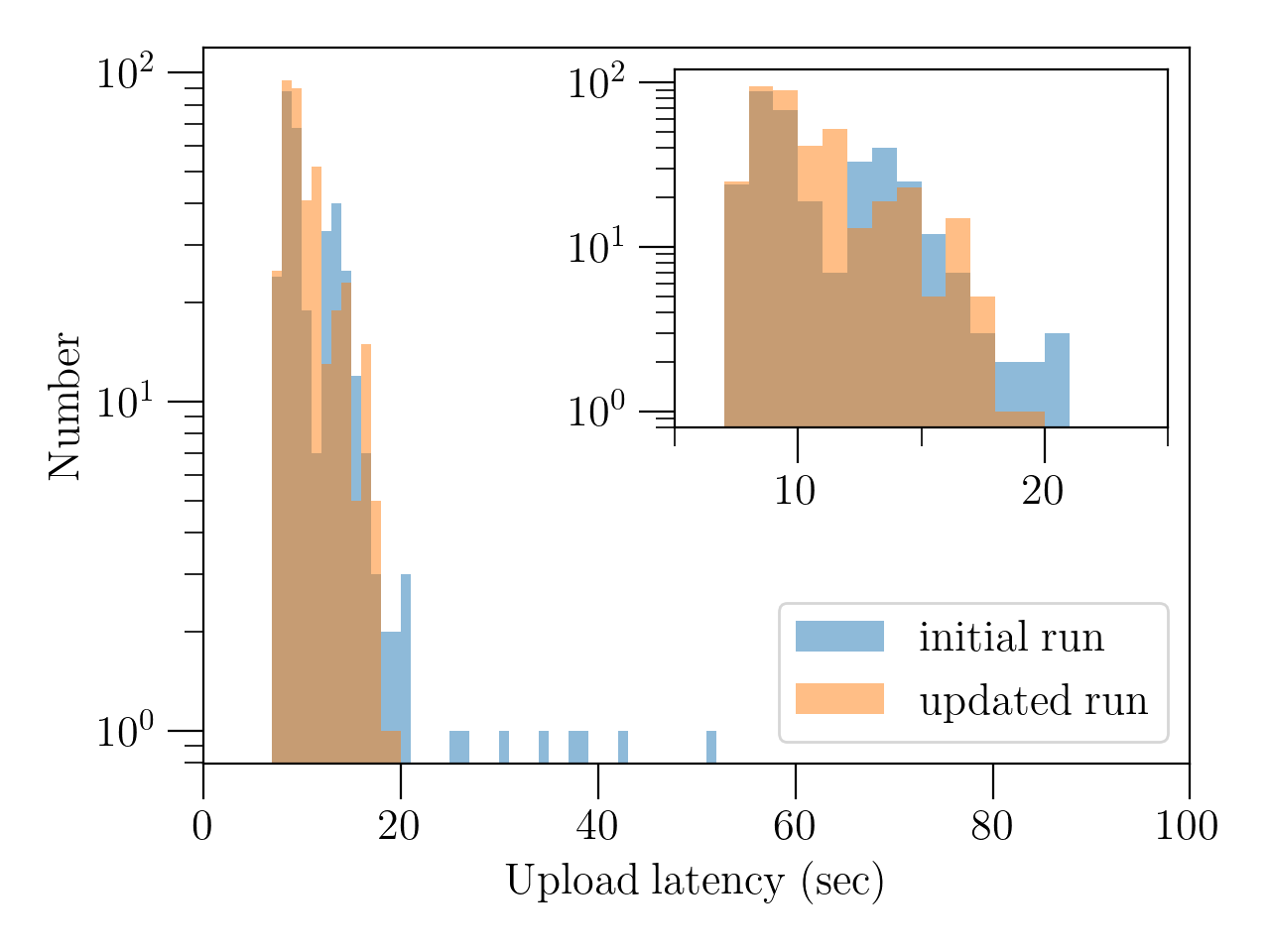}
\caption{\label{fig:latency-hist-mdc12}
Histogram of event upload latencies for the initial run (orange) and updated run (blue). 
}
\end{figure}

\begin{figure}[h]
\includegraphics[scale=0.5]{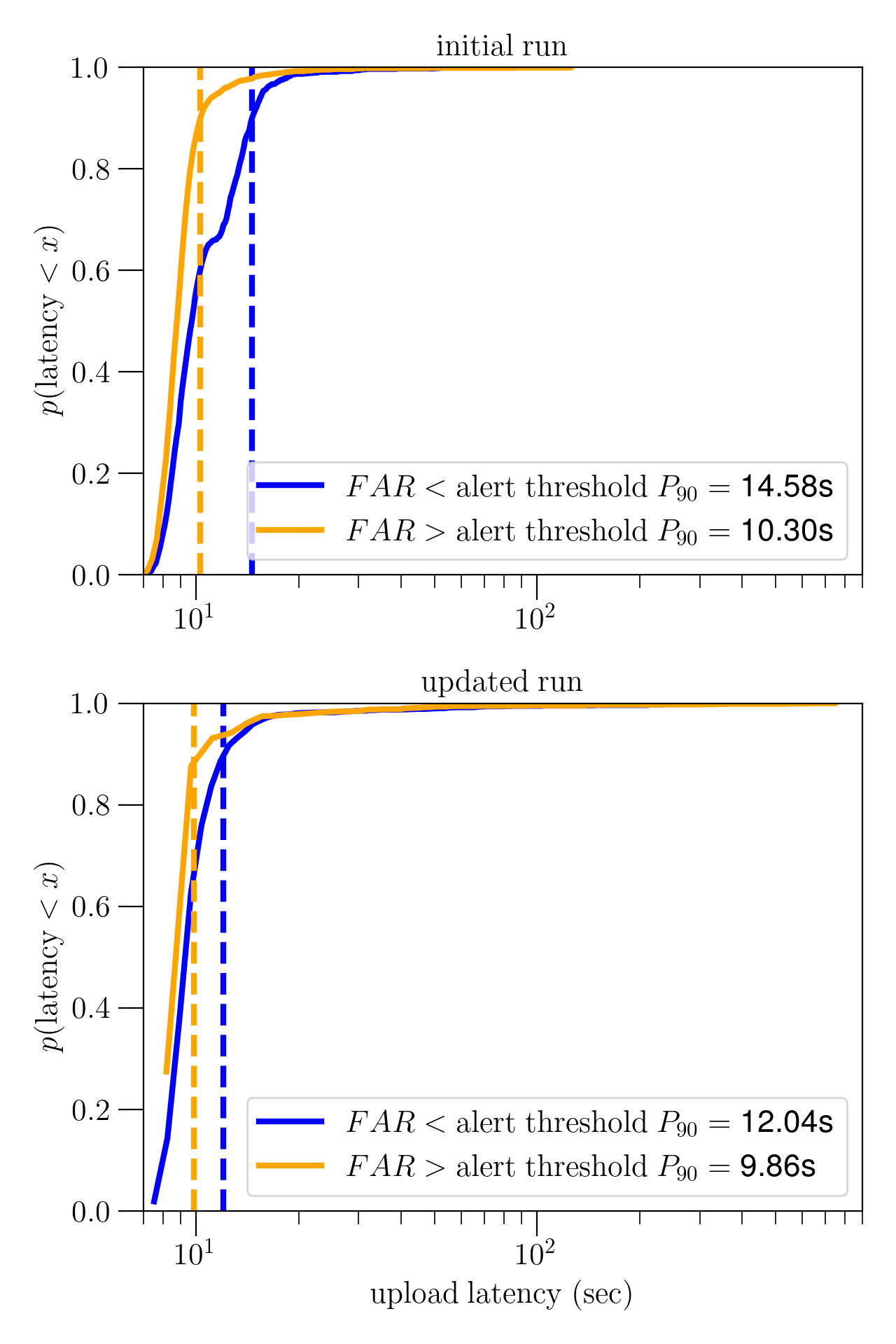}
\caption{\label{fig:latency-cdf-mdc12}
Cumulative distribution of event upload latencies for the initial run (upper panel) and updated run (lower panel). 
The orange (blue) curve shows the distribution of latencies for the first event uploaded with \ac{FAR} higher (lower) than the public alert threshold. 
The dashed lines show the location of the $90^{th}$ percentile for each distribution. 
}
\end{figure}

\figref{fig:mdc12-pastro} is a Sankey diagram showing the \PASTRO{} classification performance in the updated run. 
The initial results showed a significant amount of misclassification between \ac{BNS} and \ac{NSBH} as well as between \ac{NSBH} and \ac{BBH} injections. 
The updated run shows improvements in these areas, with only \BNSTONSBHMDCTWELVE{} of recovered \ac{BNS} being classified as \ac{NSBH} and \NSBHTOBBHMDCTWELVE{} of recovered \ac{NSBH} being classified as \ac{BBH}. The reason for this improvement is as follows.

Throughout the \PASTRO{} calculation, miss-classification of sources are assumed to be the result of Gaussian noise fluctuations causing a GW signal to match better with templates that are further away in parameter space than the one that would recover it in the absence of noise~\cite{Fong:2018,Ray:2023nhx}. 
For the initial run, the inner product of this random noise time series with itself was assumed to be  normally distributed with zero mean and unit variance~\cite{Fong:2018}. 
In the updated run, we instead make the more realistic assumption that the matched filter of the random noise time-series with the recovered template waveform is the quantity whose distribution can be modeled to be Gaussian~\cite{Ray:2023nhx}. 
Further improvements in classification are expected to result upon changing the population model in Eq.~\eqref{eq:pop-model} with the true distribution of source parameters~\cite{Ray:2023nhx}.

\begin{figure}[h]
\includegraphics[width=\columnwidth]{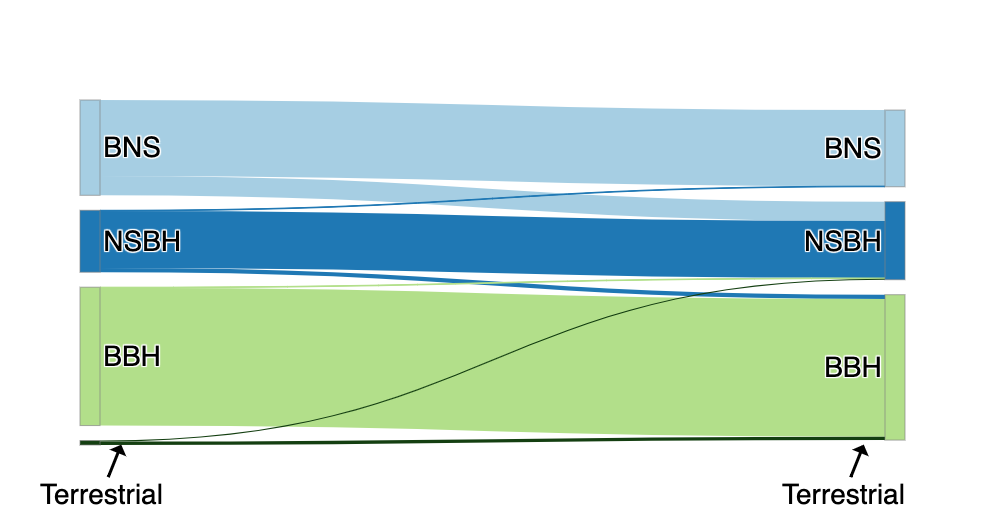}
\caption{\label{fig:mdc12-pastro}
Sankey diagram showing \PASTRO{} classification of events uploaded to \ac{GRACEDB} during the updated run from 04-24-2023 to 05-07-2023. 
Events with an end time within $\pm 1$ second of an injection are classified as either \ac{BNS}, \ac{NSBH}, or \ac{BBH} using a neutron star mass boundary of \NSMASSHIGH{}. 
Events that do not correspond in time with an injection are all classified as terrestrial.
}
\end{figure}

\end{acknowledgments}

\clearpage
\bibliography{references}

\end{document}